\documentclass[onecolumn]{IEEEtran}
\usepackage{booktabs}
\usepackage{hhline}
\usepackage{cite}
\usepackage[T1]{fontenc}
\usepackage{bbm}
\usepackage{amssymb}
\usepackage{amsfonts}
\usepackage{epsfig}
\usepackage{calc}
\usepackage{color}
\usepackage[all]{xy}
\usepackage{bm}
\usepackage{algpseudocode}
\usepackage{enumerate}
\usepackage{pdfsync}
\usepackage{framed}
\usepackage{caption}
\usepackage{amsthm}
\usepackage{blkarray}
\usepackage{subcaption}
\usepackage{enumitem}
\usepackage{hyperref}
\usepackage[linesnumbered,ruled,vlined]{algorithm2e}

\usepackage{mathrsfs}
\usepackage{caption}
\usepackage{float}
\newtheorem{defn}{Definition}

\newtheorem{thm}{Theorem}
\newtheorem{cor}{Corollary}
\newtheorem{prop}{Proposition}

\newtheorem{lem}{Lemma}
\newtheorem{conj}{Conjecture}
\newtheorem{constr}{Construction}

\usepackage[normalem]{ulem}

\newtheorem{remark}{Remark}
\newtheorem{example}{Example}

\newcounter{definition}[section]

\newcommand{\bit}{\begin{itemize}}
	\newcommand{\eit}{\end{itemize}}
\newcommand{\bcor}{\begin{cor}}
	\newcommand{\ecor}{\end{cor}}
\newcommand{\beq}{\begin{equation}}
	\newcommand{\eeq}{\end{equation}}
\newcommand{\beqn}{\begin{equation*}}
	\newcommand{\eeqn}{\end{equation*}}
\newcommand{\bea}{\begin{eqnarray}}
	\newcommand{\eea}{\end{eqnarray}}
\newcommand{\bean}{\begin{eqnarray*}}
	\newcommand{\eean}{\end{eqnarray*}}
\newcommand{\ben}{\begin{enumerate}}
	\newcommand{\een}{\end{enumerate}}
\newcommand{\bdefn}{\begin{defn}}
	\newcommand{\edefn}{\end{defn}}
\newcommand{\bnote}{\begin{remark}}
	\newcommand{\enote}{\end{remark}}
\newcommand{\bprop}{\begin{prop}}
	\newcommand{\eprop}{\end{prop}}
\newcommand{\blem}{\begin{lem}}
	\newcommand{\elem}{\end{lem}}
\newcommand{\bthm}{\begin{thm}}
	\newcommand{\ethm}{\end{thm}}
\newcommand{\bconj}{\begin{conj}}
	\newcommand{\econj}{\end{conj}}
\newcommand{\bconstr}{\begin{constr}}
	\newcommand{\econstr}{\end{constr}}
\newcommand{\bpf}{\begin{proof}}
	\newcommand{\epf}{\end{proof}}
\newcommand{\bprf}{{\em Proof: }}

\newcommand{\eproof}{\hfill $\Box$}
\usepackage[utf8]{inputenc} 
\usepackage[T1]{fontenc}
\usepackage{url}              
\usepackage{cite}             

\usepackage[cmex10]{amsmath}  
\usepackage{mleftright}       
\mleftright                   

\usepackage{graphicx}         
\usepackage{booktabs}         

\begin{document}

\title{Explicit Information-Debt-Optimal Streaming Codes With Small Memory} 

\author{%
  \IEEEauthorblockN{Anonymous Authors}
  \IEEEauthorblockA{%
    Please do NOT provide authors' names and affiliations\\
    in the paper submitted for review, but keep this placeholder.\\
    ISIT23 follows a \textbf{double-blind reviewing policy}.}
}

\author{ 
	\IEEEauthorblockN{	
				M. Nikhil Krishnan\IEEEauthorrefmark{1}, 
						Myna Vajha\IEEEauthorrefmark{2},
		Vinayak Ramkumar\IEEEauthorrefmark{3},
		P. Vijay Kumar\IEEEauthorrefmark{4}\\}
	\IEEEauthorblockA{\IEEEauthorrefmark{1}%
		Department of Electrical Engineering, IIT Palakkad\\}
	\IEEEauthorblockA{\IEEEauthorrefmark{2}%
		Qualcomm, Bengaluru\\}
	\IEEEauthorblockA{\IEEEauthorrefmark{3}	Department of Electrical Engineering--Systems, Tel Aviv University\\} 
		\IEEEauthorblockA{\IEEEauthorrefmark{4}%
		Department of Electrical Communication Engineering, IISc Bangalore\\}
	\IEEEauthorblockA{
		\{nikhilkrishnan.m, mynaramana, vinram93, pvk1729\}@gmail.com}
	\thanks{The work of M. Nikhil Krishnan is supported by a faculty seed grant from the Indian Institute of Technology Palakkad and the DST-INSPIRE faculty fellowship. The work of P. Vijay Kumar is supported by the ``Next Generation Wireless Research and Standardization on 5G and Beyond'' project funded by MeITY and the SERB Grant No.~CRG/2021/008479.
	} 
}
\IEEEoverridecommandlockouts
\maketitle

\begin{abstract}
	For a convolutional code in the presence of a symbol erasure channel, the information debt $I(t)$ at time $t$ provides a measure of the number of additional code symbols required to recover all message symbols up to time $t$. Information-debt-optimal streaming ($i$DOS) codes are convolutional codes which allow for the recovery of all message symbols up to $t$ whenever $I(t)$ turns zero under the following conditions; (i) information debt can be non-zero for at most $\tau$ consecutive time slots and (ii) information debt never increases beyond a particular threshold. The existence of periodically-time-varying $i$DOS codes are known for all parameters. In this paper, we address the problem of constructing explicit, time-invariant $i$DOS codes. We present an explicit time-invariant construction of $i$DOS codes for the unit memory ($m=1$) case.  It is also shown that a construction method for convolutional codes due to Almeida et al. leads to explicit time-invariant $i$DOS codes for all parameters. However, this general construction requires a larger field size than the first construction for the $m=1$ case.    
\end{abstract}

\section{Introduction\label{sec:intro}}

Streaming codes are convolutional codes that ensure decoding within a worst-case delay.  In the streaming code literature \cite{MartSunTIT04,MartTrotISIT07,BadrKTAInfocom13,BadrPatilKhistiTIT17,NikPVK,simple,FongKhistiTIT19,NikDeepPVK,RamVajNikPVK_SDE,KhistiExplicitCode,VajRamNikPVK,RamVajPVK_generalized_simple,RamVajPVK_locally_recoverable_SC, Rashmi_Variable_packet_size}, packet-erasure channel models are often considered. In contrast, we focus on codes over a more general symbol-erasure channel model in this work. 
Let $(n,k,m)$ be the parameters of a convolutional code, where $k$ is the number of message symbols per time slot, $n > k$ is the number of code symbols per time slot, and $m$ is the memory.  The  information debt is a measure of the number of additional coded symbols needed to decode the message symbols encoded thus far. This notion was first introduced by Martinian  in \cite{Martinianthesis}. The error probability of random linear streaming codes in the large field size regime over i.i.d. symbol erasure channels is characterized  in \cite{SuHuaLinWanWan}  using an information-debt-based argument. 

  Consider symbol erasure patterns such that information debt stays positive for no more than $\tau$ consecutive time slots.
  If the code is capable of  recovering message symbols whenever information debt drops to zero,  then $\tau$ can be thought of as a worst-case decoding delay. It is argued in \cite{idos} that if information debt goes above $mk$ in any time slot, then it is not possible to recover all message symbols. With these in mind, the authors of \cite{idos} defined $(n,k,m,\tau)$ $i$DOS codes as  an $(n,k,m)$ convolutional code that is capable of decoding all previously unknown messages in any time slot where information debt becomes zero provided the  symbol erasure pattern is such that i) information debt does not stay positive for more than $\tau$ successive time slots and ii) information debt never crosses $mk$. 
   
  For $\tau \le m$, the existence of periodically-time-varying $(n,k,m,\tau)$ $i$DOS codes over a sufficiently large field follows from the results in \cite{Martinianthesis}.
    In \cite{idos}, this result is extended to all valid parameters over $\mathbb{F}_q$, with $q>(\tau+1){n(\tau+1) \choose k(\tau+1)}.$
These existence results are based on Combinatorial Nullstellensatz \cite{Alon_Comb} and  hence provide no insights that will lead to an explicit construction. The connection of $i$DOS codes with two well-known classes of (time-invariant) convolutional codes, namely $m$-MDS codes \cite{Gabidulin} and maximum distance profile (MDP) codes \cite{MDS_conv},  is established in \cite{idos}. For $\tau \le m$, $m$-MDS codes are shown to be $i$DOS codes, and for $\tau \le m+ \lceil \frac{mk}{n-k}\rceil $, MDP codes are shown to be $i$DOS codes. For parameters $\{n,k=1,m=1,\tau\}$, a special case of the MDP code in \cite{MDP_RS}  is shown to yield an explicit construction of $i$DOS codes over a field of size $O(n)$.  Apart from the results mentioned above, the paper \cite{idos} does not provide explicit $i$DOS codes for $m<\tau$. Small memory is advantageous in scenarios where low complexity encoders are required, such as in sensor networks. Furthermore, having a larger $\tau$ for a given $m$ implies recoverability from a larger set of erasure patterns. The question of whether time-invariant $i$DOS codes always exist for $m\ll\tau$ is also unanswered in \cite{idos}.
\subsection*{Our Contributions}
\begin{itemize}[leftmargin=*]
\item  As the primary result, we provide an explicit, time-invariant construction of $i$DOS codes over $\mathbb{F}_{2^d}$ with unit memory $(m=1)$, for all valid $\{n,k,\tau\}$, where $d>(n-1)k^2(\tau+1)$.   
\item We also show that an explicit, time-invariant construction of $i$DOS codes for all possible $\{n,k,m,\tau\}$ follows from a convolutional code construction method due to Almeida et al. \cite{AlmNapPin}.  This construction is over a finite field of size $2^d$, where $d=O(2^{{\tiny }mn+k}(\tau+1)k)$. Notably, for the unit memory case, the former construction requires a smaller field size. 

\eit 

\paragraph*{Organization of the Paper}

 In Section~\ref{sec:prelim}, we first define $i$DOS codes. Then, we  present some  definitions and known results that are needed for the later sections. Our unit memory construction is presented in Section~\ref{sec:unit_memory_construction}. In  Section~\ref{sec:general_construction}, we present the general construction.

\paragraph*{Notation}
We use $\mathbb{N}$ to denote $\{1,2,3,\dots\}$. If $r<s$, we will interpret the sum $\sum_{i=s}^{r}x_i$ as being equal to $0$.  
For integers $x,y$, we define the set of integers $[x:y]=\{i\mid x\leq i\leq y\}$.  Furthermore, we use the notation $[x]$ to denote $[1:x]$. We use $\mathbb{F}_q$ to denote the finite field consisting of $q$ elements, where $q$ is a prime power. For an $x\times y$ matrix $A$ and $\mathcal{S}\subseteq [x]$, let $A(\mathcal{S},:)$ denote the submatrix obtained by restricting $A$ to the rows in $\mathcal{S}$. Similarly, for $\mathcal{T}\subseteq [y]$, we use $A(:,\mathcal{T})$ to denote the submatrix obtained by restricting $A$ to the columns in $\mathcal{T}$. Moreover, $A(\mathcal{S},\mathcal{T})$ denotes the $|\mathcal{S}|\times|\mathcal{T}|$ submatrix obtained by restricting  $A(\mathcal{S},:)$ to the columns in $\mathcal{T}$. We use $A(i,j)$ to denote the element in row-$i$ and column-$j$ of $A$.  Let $M$ denote an $x\times y$ matrix whose entries are drawn from $\{-\infty,0\}\cup\mathbb{N}$. If $\alpha\in \mathbb{F}_q$, then $\alpha^M$ denotes an $x\times y$ matrix over $\mathbb{F}_q$ whose entries are given by $\{\alpha^{M(i,j)}\mid 1\leq i\leq x, 1\leq j\leq y\}$. Here, we set $\alpha^{-\infty}\triangleq 0$. We use $\deg(f(x))$ to denote the degree of a polynomial $f(x)$. 
For integer $x\geq 1$, let $S_x$ denote the symmetric group consisting of all permutations of the set $[x]$. Let $\mathcal{S}\subseteq [x]$ and $\sigma\in S_x$. We define $\sigma(\mathcal{S})\triangleq \{\sigma(i)\mid i\in \mathcal{S}\}$.

\section{Preliminaries} \label{sec:prelim}
The first three subsections of this section focus on providing background on the $i$DOS code setting, for which we follow the notation from \cite{idos}. The latter part of the section introduces some definitions and results that are needed for the proofs of explicit constructions.
\subsection{Convolutional Codes}\label{sec:conv_codes}
An $(n,k,m)$ convolutional code over $\mathbb{F}_q$ can be described as follows. The encoder gets $k$ message symbols and outputs $n>k$ coded symbols in each time slot $t \in \mathbb{N}$.  These message symbols are denoted by $\underline{s}(t)=[s_1(t) \dots s_k(t)]^T \in \mathbb{F}_q^k$ and the code symbols are given by $\underline{c}(t)=[c_1(t) \dots c_n(t)]^T \in \mathbb{F}_q^n.$ The memory of the encoder is $m$.  There is an $n \times (m+1)k$ matrix over $\mathbb{F}_q$, denoted by $G_t$, such that 
\bean 
\underline{c}(t)=G_t \begin{bmatrix}
	\underline{s}(t-m)\\ \underline{s}(t-m+1) \\ \dots \\ \underline{s}(t)
\end{bmatrix}. 
\eean   
For all $t \le 0$, we set $\underline{s}(t)=\underline{0}$.  The convolutional code is said to be time-invariant if $G_{t}$ is the same for all $t \in \mathbb{N}$.  Let  $G=[G^{(m)} ~ G^{(m-1)}~\dots~G^{(0)}],$ where each $G^{(i)}$ is an $n \times k$ matrix over $\mathbb{F}_q$. The current paper focuses only on time-invariant constructions, and hence we set $G_t=G$ for all $t \in\mathbb{N}$. By abuse of notation, we will refer to $G$ as the {\it generator matrix}.  The $n$ symbols belonging to $\underline{c}(t)$ are sent to the receiver in time slot $t$.  We describe symbol erasure patterns using sets $\mathcal{R}_t \subseteq [n]$  such that the receiver receives  $\{c_j(t) \mid j \in \mathcal{R}_t\}$ in time slot $t$ and $\{c_j(t) \mid j \in [n] \setminus \mathcal{R}_t\}$ are erased by the channel. The number of non-erased code symbols in time slot $t$ is denoted by  $n_t=|\mathcal{R}_t|$. 


\subsection{Information Debt}\label{sec:info_debt}
Information debt, introduced by Martinian \cite{Martinianthesis}, is a measure of the extra code symbols needed at the decoder to decode all unknown message symbols.  The information debt at time slot $0$ is set as zero. The information debt at any time slot $t \in \mathbb{N}$ is given by $I(t)=\max\{k-n_t+ I(t-1),0\}~\text{for all}~t \in \mathbb{N}.$ 
Let $\theta_0=0$. For any symbol erasure pattern, one can identify time slots $\{\theta_i\}_{i=1}^{\infty}$ such that $$\theta_{i+1}=\inf \{t>\theta_{i} \mid I(t) = 0\}.$$ 
 
\subsection{$i$DOS Codes}
The goal of $i$DOS codes is to recover messages whenever information debt drops to zero.  It is shown in \cite{idos} that if $I(t)>mk$ for any time slot $t$, then there exists no $(n,k,m)$ convolutional code capable of decoding all the message symbols. To ensure a worst-case decoding delay of  $\tau$ whenever recovery is possible, it is required that $\theta_{i+1}-\theta_{i} \le \tau+1$, for all $i$. 
With this background, $i$DOS codes can be  formally defined as follows. 
\begin{defn}[\cite{idos}]\label{def:idos_code}\normalfont A symbol erasure pattern is said to be $(n,k,m,\tau)$-acceptable if   $I(t) \le mk$ for all $t \in \mathbb{N}$,  and 
$\theta_{i+1}-\theta_{i} \le \tau+1$ for all $i \in \mathbb{N} \cup \{0\}$.
An $(n,k,m,\tau)$ $i$DOS code is an $(n,k,m)$ convolutional code which is such that for all $i \in \mathbb{N} \cup \{0\}$, $\{\underline{s}(t) \mid t \in [\theta_{i}+1:\theta_{i+1}]\}$ are recoverable at time $\theta_{i+1}$  over every $(n,k,m,\tau)$-acceptable symbol erasure pattern. 
\end{defn}

The requirement that the messages need to be recovered whenever information debt drops to zero, is intuitively related to the non-singularity of certain matrices obtained from the generator matrix. In the rest of this section, we will identify a sufficient property to be possessed by an $x\times x$ matrix $M$ so that $\alpha^M$ is non-singular. This idea forms the core of our unit-memory construction.
	

\subsection{Dominant Permutation}
\begin{defn}[Dominant Permutation of a Matrix]\normalfont Consider an  $x\times x$ matrix $M$ composed of elements drawn from $\{-\infty,0\}\cup\mathbb{N}$. The permutation $\sigma^*\in S_x$ (if exists) is referred to as the {\it dominant permutation} if the following is true:
	$$\sum_{i=1}^x{M({\sigma^*(i),i})}> \sum_{i=1}^x{M({\sigma(i),i})}, \forall \sigma\in S_x\setminus\{\sigma^*\}.$$
\end{defn}
Furthermore, if such a $\sigma^*$ exists, we refer to the sum $\sum_{i=1}^x{M({\sigma^*(i),i})}$ as the {\it dominant sum} of $M$.  We make the following simple observation. 

\begin{remark}\label{rem:infty_remark}\normalfont
	Assume that for the matrix $M$, we have $M({i,j})=-\infty$. If $\sigma\in S_x$ is such that $\sigma(j)=i$, clearly, $\sigma$ cannot be a dominant permutation of $M$.  
\end{remark}

\begin{defn}[Dominant Submatrix]\normalfont
	Consider an $x\times y$ matrix $M$ composed of elements from $\{-\infty,0\}\cup\mathbb{N}$, where $y\leq x$. A $y\times y$ submatrix  $\tilde{M}$ of $M$ (if exists) is referred to as the {\it dominant submatrix} of $M$, if the following two conditions hold; (i) $\tilde{M}$ possesses a dominant permutation, (ii) among all the $y\times y$ submatrices which possess a dominant permutation, $\tilde{M}$ yields the single largest dominant sum -- i.e., dominant sums (if exist) of all other submatrices are {\it strictly} smaller. 
\end{defn}
We make the following straightforward observation which relates dominant submatrices and the existence of dominant permutation.
\begin{remark}[Submatrix Decomposition Strategy]\label{rem:dom_sub_matrix}\normalfont
	Consider an $x\times x$ matrix $M$ composed of elements from $\{-\infty,0\}\cup\mathbb{N}$. Let $\mathcal{A}_1,\ldots,\mathcal{A}_l$ denote a partition of the columns $[x]$ of $M$. Assume that for each $\mathcal{A}_i$, the submatrix $M(:,\mathcal{A}_i)$ possesses a dominant submatrix ${M}_{\mathcal{A}_i}$. Let ${M}_{\mathcal{A}_i}$ be occupying the rows $\mathcal{B}_i\subseteq [x]$ and let $s_i$ denote the dominant sum of ${M}_{\mathcal{A}_i}$. If the rows $\mathcal{B}_1,\ldots,\mathcal{B}_l$ do not intersect (i.e., they form a partition of $[x]$), it follows that $M$ possesses a dominant permutation. Moreover, the dominant sum of $M$ is given by $\sum_{i=1}^l s_i$. We will refer to this simple strategy of showing the existence of the dominant permutation for a larger matrix by leveraging the existence of smaller dominant submatrices whose rows do not intersect, as the {\it submatrix decomposition strategy}.
\end{remark}

\begin{figure}[!]
	\begin{center}
		\includegraphics[scale=0.25]{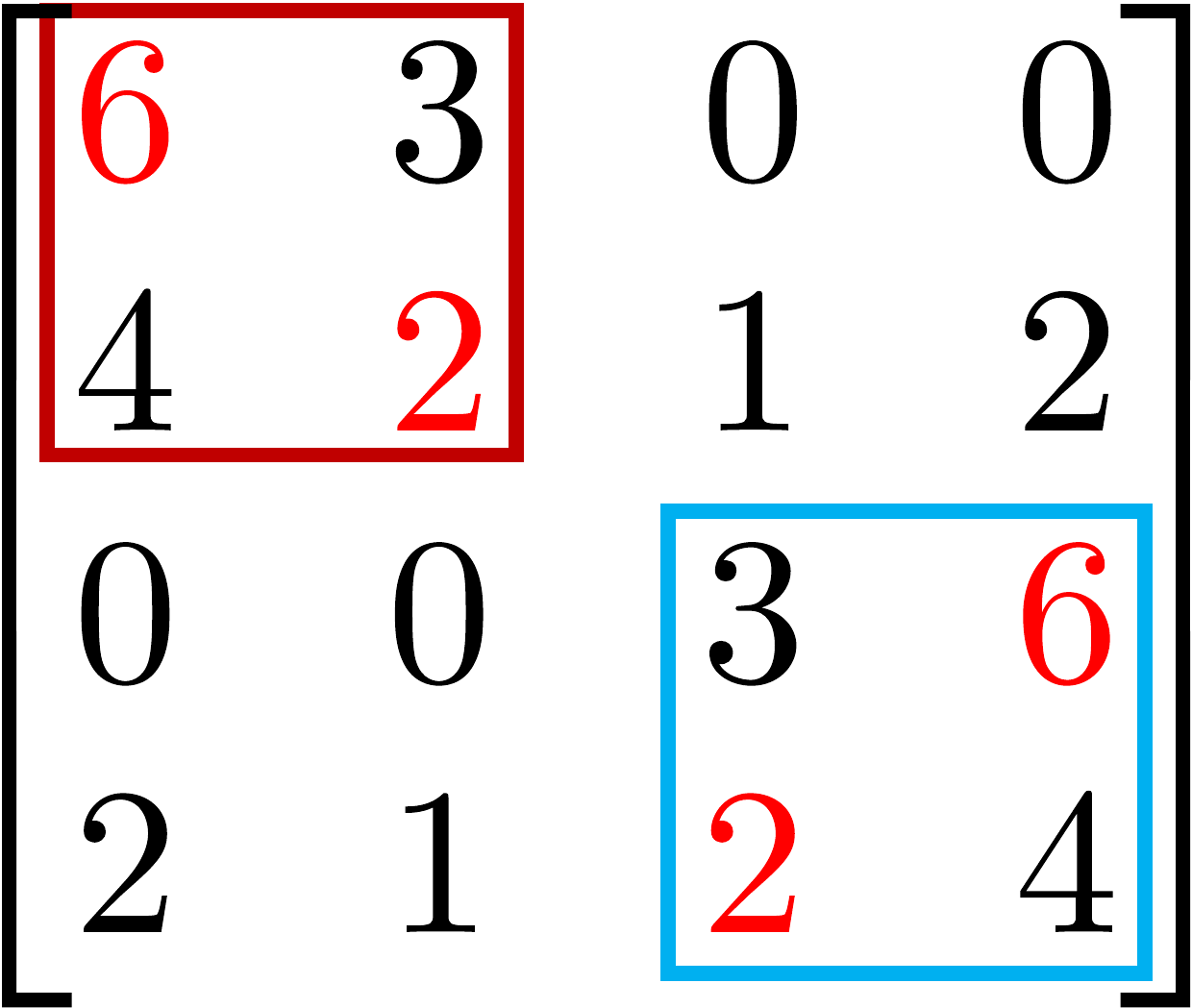}
		\caption{In this figure, we illustrate how the submatrix decomposition strategy can be utilized to show that the given matrix $M$ possesses a dominant permutation. We choose: $\mathcal{A}_1=\{1,2\}, \mathcal{A}_2=\{3,4\}$. The  dominant submatrices associated with $M(:,\mathcal{A}_1)$ and $M(:,\mathcal{A}_2)$ are demarcated using red and blue rectangles, respectively. As the rows of these submatrices do not intersect, it follows that $M$ possesses a dominant permutation.}
		\label{fig:dominant_perm}
	\end{center}
\vspace{-0.15in}
\end{figure} 

For the matrix illustrated in Fig.~\ref{fig:dominant_perm}, the dominant permutation is given by $\sigma^*=(1,2,4,3)$ and the corresponding dominant sum is $16$. In the following example, we discuss a slight variation of the submatrix decomposition strategy by making use of the observation made in Remark \ref{rem:infty_remark}.

\begin{example}\normalfont\label{ex:constrained_dom}
	Consider the matrix $M$ illustrated in Fig.~\ref{fig:dominant_perm_2}. Let $\mathcal{A}_1=\{1,2\}$ and $\mathcal{A}_2=\{3,4\}$. We highlight the corresponding dominant submatrices using green and blue rectangles, respectively. As the rows of these submatrices intersect, we cannot employ the naive submatrix decomposition strategy. However, from Remark \ref{rem:infty_remark}, it can be inferred that as $M({1,3})=M({1,4})=-\infty$, if there is a dominant permutation $\sigma^*$, it should be that $1\notin\sigma^*(\mathcal{A}_2)$. In other words, $1\in \sigma^*(\mathcal{A}_1)$. Essentially, the implication here is that $\sigma^*$ should ``pass through'' row $1$, when restricted to the columns in $\mathcal{A}_1$. As a result, while searching for the dominant submatrix of $M(:,\mathcal{A}_1)$, we will consider only those submatrices which involve row $1$. Given this constraint, it can be identified that the submatrix demarcated using the red dashed rectangle is the (constrained) dominant submatrix. Since the rows of the constrained dominant submatrix and the dominant submatrix (indicated in blue) are not intersecting, it follows that the matrix $M$ possesses a dominant permutation. We will utilize this idea of constrained dominant submatrices later in our proofs.  
	
	\begin{figure}[!]
		\begin{center}
			\includegraphics[scale=0.25]{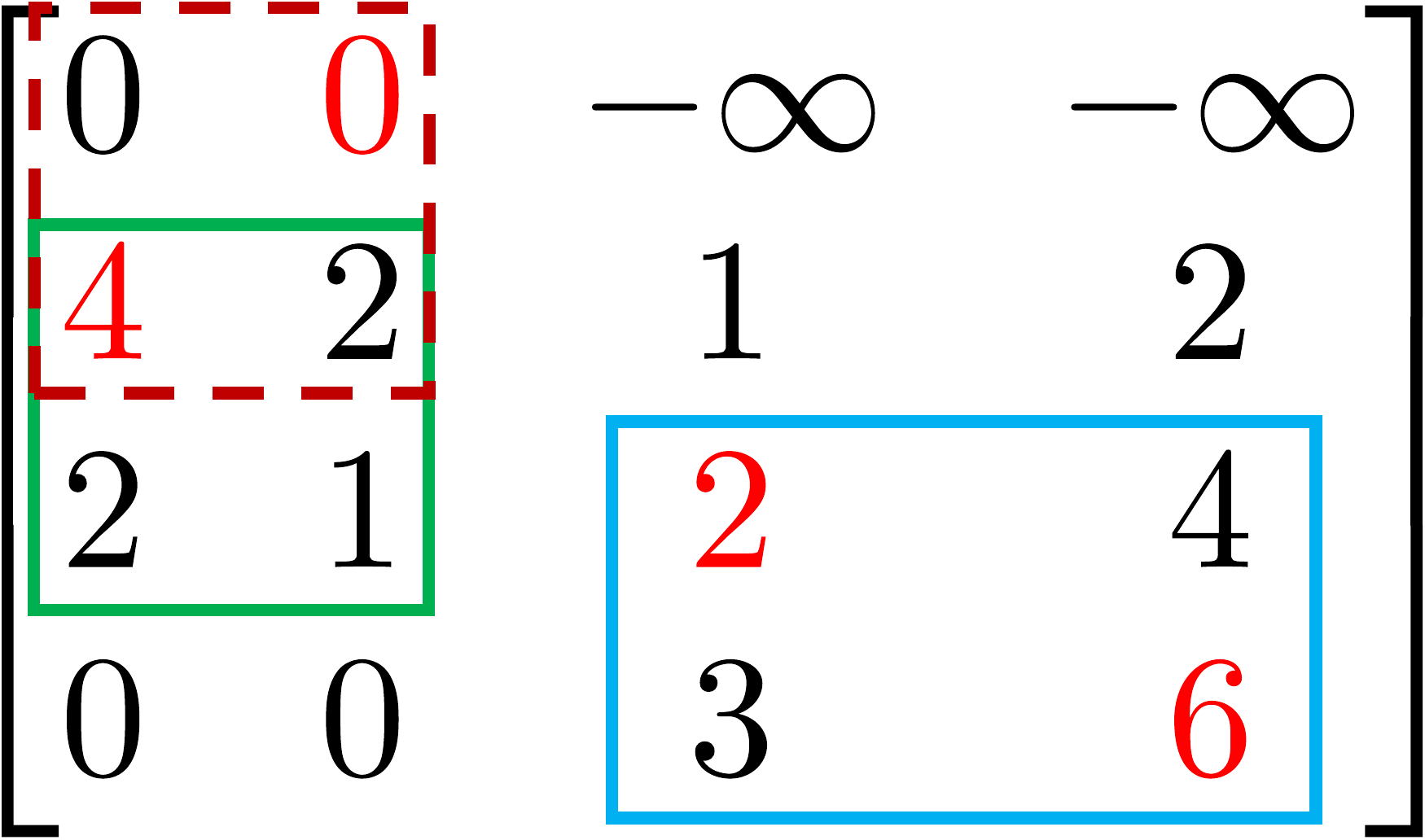}
			\caption{For the given matrix $M$, the dominant permutation is $\sigma^*=(2,1,3,4)$. We argue the existence of the dominant permutation in Example \ref{ex:constrained_dom} using a variant of the submatrix decomposition strategy.}
			\label{fig:dominant_perm_2}
		\end{center}
	\end{figure} 
	
\end{example}

The following lemma motivates our ongoing discussion on matrices that possess dominant permutations. We attribute this lemma to an earlier work by Almeida et al. \cite{AlmNapPin}, which explores similar ideas. 
 The proof of this lemma is deferred to Appendix~\ref{sec:dominant_perm_non_singularity_proof}.


\begin{lem}\label{lem:dominant_perm_non_singularity}
	Consider an $x\times x$ matrix $M$ with elements drawn from $\{-\infty,0\}\cup\mathbb{N}$. Assume $M$ possesses the dominant permutation $\sigma^*$ and the dominant sum $s_{\sigma^*}$. Let $\alpha$ be a primitive element of  $\mathbb{F}_{p^d}$, where $d>s_{\sigma^*}$. Then, $\alpha^M$ is non-singular.
\end{lem}

\section{Construction A: Explicit, Unit Memory Construction} \label{sec:unit_memory_construction}
In this section, we present an explicit construction of $i$DOS codes with unit memory (i.e., $m=1$), for all parameters $\{n,k,\tau\}$. Without loss of generality, we will henceforth take the characteristic of the underlying finite field for our constructions to be two, i.e., $p=2$. To highlight the key ideas, we first discuss an example for parameters $\{n=4,k=2,\tau=2\}$ in Sec.~\ref{sec:example}. The general construction for any $\{n,k,\tau\}$ is described in Sec.~\ref{sec:explicit_unit_mem_constr}. 

\subsection{Example: $\{n=4,k=2,m=1,\tau=2\}$}\label{sec:example}


Let $\alpha$ be a primitive element of $\mathbb{F}_{2^d}$. For now, we will assume $d$ to be sufficiently large and later in the section we will explicitly specify a value for $d$. We set $G^{(0)}\triangleq \alpha^{M^{(0)}}$ and $G^{(1)}\triangleq \alpha^{M^{(1)}}$, where: $$M^{(0)}\triangleq
\def\r#1{\color{red}{#1}}
\begin{bmatrix} \color{red} \r0 & \r0 \\ \r1 & \r2\\ \r2& \r4\\ \r3 & \r6 \end{bmatrix},~~ M^{(1)}\triangleq \def\b#1{\color{blue}{#1}} \begin{bmatrix} \b6 & \b3 \\ \b4 & \b2\\ \b2& \b1\\ \b0 & \b0 \end{bmatrix}.$$ 

Recall the definitions of $\mathcal{R}_t$, $\{\theta_i\}$ presented in Sections \ref{sec:conv_codes} and \ref{sec:info_debt}, respectively. Let $\theta_{i+1}-\theta_i\triangleq \ell$. We will now show that all the message symbols in $\{\underline{s}(t)\mid t\in [\theta_i+1:\theta_{i}+\ell]\}$ can be recovered by the receiver using the available non-erased code symbols $\{c_j(t)\mid t\in [\theta_i+1:\theta_{i}+\ell],j\in \mathcal{R}_t\}$. This is under the assumption that the message symbols in $\{\underline{s}(t')\mid t'\in [\theta_i]\}$ are already known to the receiver (the assumption is trivially true when $i=0$). After removing the contribution of these message symbols, it is as if the transmitter has sent $\underline{\hat{c}}(\theta_i+1)=G^{(0)}\underline{s}(\theta_i+1)$ and $\underline{\hat{c}}(t')=G^{(1)}\underline{s}(t'-1)+G^{(0)}\underline{s}(t')$, where $t'\in [\theta_i+2:\theta_i+\ell]$. Thus, effectively, the received code symbols in time slots $[\theta_i+1:\theta_i+\ell]$ are given by
$\{\hat{c}_j(t)\mid t\in [\theta_i+1:\theta_{i}+\ell],j\in \mathcal{R}_t\}$, where $\underline{\hat{c}}(t)=[\hat{c}_1(t)~\cdots~\hat{c}_n(t)]^T$. 

By definition of $\{\theta_i\}$, we have $I(\theta_i)=I(\theta_i+\ell)=0$. Hence, $\sum_{t\in[\theta_i+1:\theta_i+\ell]}n_t\geq k\ell= 2\ell$. Without loss of generality, we will consider here only the worst-case scenario $\sum_{t\in[\theta_i+1:\theta_i+\ell]}n_t= 2\ell$. In addition, by Definition \ref{def:idos_code}, we have $I(t)\leq mk=2$ and $\ell\leq \tau+1=3$. As $I(t)\leq 2$ and $I(t)>0$ for $t\in [\theta_i+1:\theta_i+\ell-1]$, it follows that: $2(\ell'-1)\leq \sum_{t\in[\theta_i+1:\theta_i+\ell']}n_t <2\ell'$, for all $\ell'\in [\ell-1]$. Thus, we restrict ourselves to $\{n_{\theta_i+1},\ldots,n_{\theta_i+\ell}\}$ satisfying three conditions: 
\bit
\item[(1)] $\ell\leq \tau+1=3$,
\item[(2)] $\sum_{t\in[\theta_i+1:\theta_i+\ell]}n_t= 2\ell$,
\item[(3)] $2(\ell'-1)\leq \sum_{t\in[\theta_i+1:\theta_i+\ell']}n_t <2\ell'$, $\ell'\in[\ell-1]$.
\eit

Let $\mathcal{R}_t=\{i_1,i_2,\ldots,i_{n_t}\}\subseteq[n]$ and $\underline{\tilde{c}}(t)\triangleq [\hat{c}_{i_1}(t)~\cdots~\hat{c}_{i_{n_t}}(t)]^T$. At time $(\theta_i+\ell)$, thus the decoder essentially has the following matrix equation to solve:
\beqn
\begin{bmatrix}
	\underline{\tilde{c}}(\theta_i+1)\\ \vdots \\ \underline{\tilde{c}}(\theta_i+\ell)
\end{bmatrix} =
G_\text{dec}\begin{bmatrix}
	\underline{s}(\theta_i+1)\\ \vdots \\ \underline{s}(\theta_i+\ell)
\end{bmatrix},
\eeqn
where $G_\text{dec}$ is a {\it decoding matrix} of size $k\ell \times k\ell$ (i.e., $2\ell\times 2\ell$). Note that $G_\text{dec}$ is not a constant and is a function of the symbol erasure pattern. The code is an $i$DOS code if and only if $G_\text{dec}$ is non-singular for any symbol erasure pattern such that  $\{n_{\theta_i+1},\ldots,n_{\theta_i+\ell}\}$ satisfy the conditions (1)--(3). The structure of decoding matrices for all the possible symbol erasure scenarios is illustrated in Fig.~\ref{fig:example_ell_1_2_3}. We illustrate the corresponding exponents (with respect to $\alpha$) in Fig.~\ref{fig:example_ell_1_2_3_exp}. The high-level idea of the construction is the following. For each $2\ell \times 2\ell$ decoding matrix $G_\text{dec}$, we have a corresponding $2\ell \times 2\ell$ {\it exponent matrix} $M_\text{dec}$ in Fig.~\ref{fig:example_ell_1_2_3_exp}. We will show that all the exponent matrices have dominant permutations. This will prove that for a large enough degree of the field extension $d$, the corresponding decoding matrices are non-singular matrices (by applying Lemma \ref{lem:dominant_perm_non_singularity}). We will specify an explicit value for $d$ later in the section.

\begin{figure*}[!]
	\begin{center}
		\includegraphics[scale=0.45]{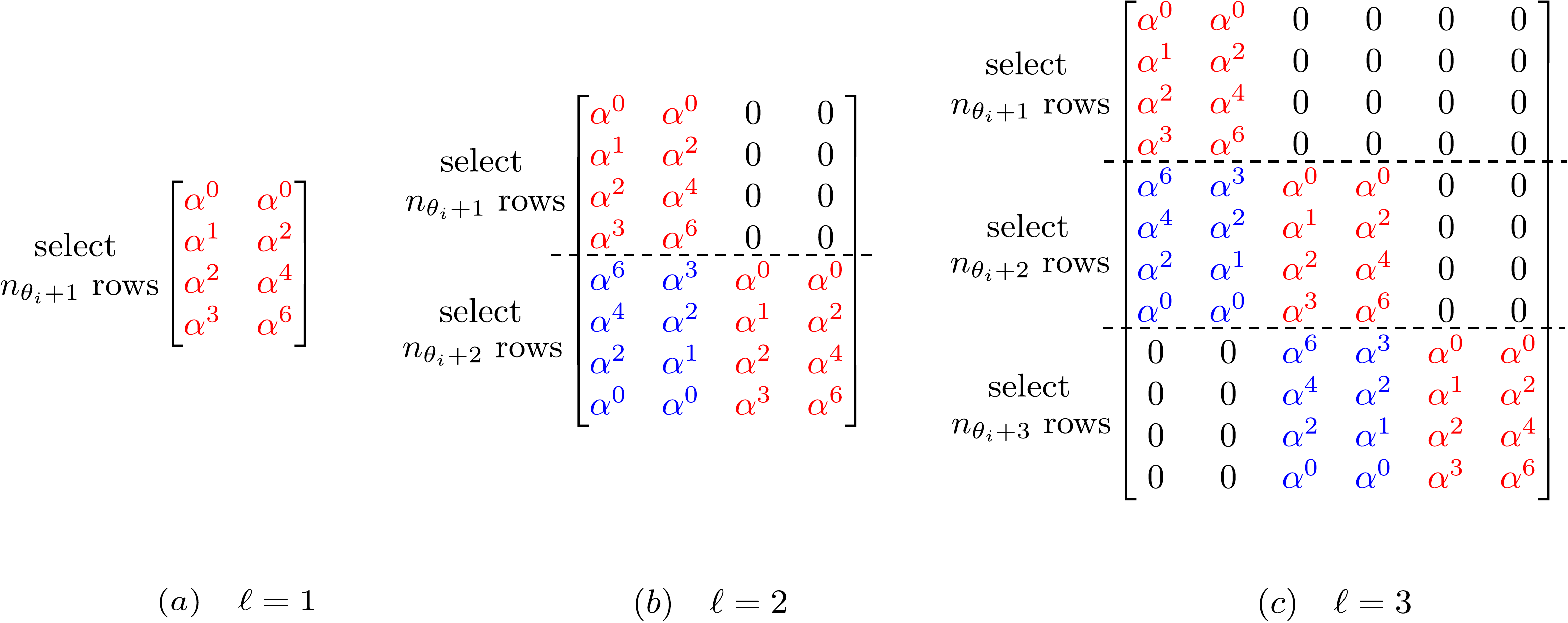}
		\caption{Let $\{n_{\theta_i+1},\ldots,n_{\theta_i+\ell}\}$ be such that they satisfy conditions (1)--(3) listed in Sec.~\ref{sec:example}. The code in the example is an $(n=4,k=2,m=1,\tau=2)$ $i$DOS code if and only if the $2\ell\times 2\ell$ decoding matrix $G_\text{dec}$ obtained by selecting rows as indicated above is non-singular. }
		\label{fig:example_ell_1_2_3}
	\end{center}
\end{figure*} 

\begin{figure*}[!]
	\begin{center}
		\includegraphics[scale=0.45]{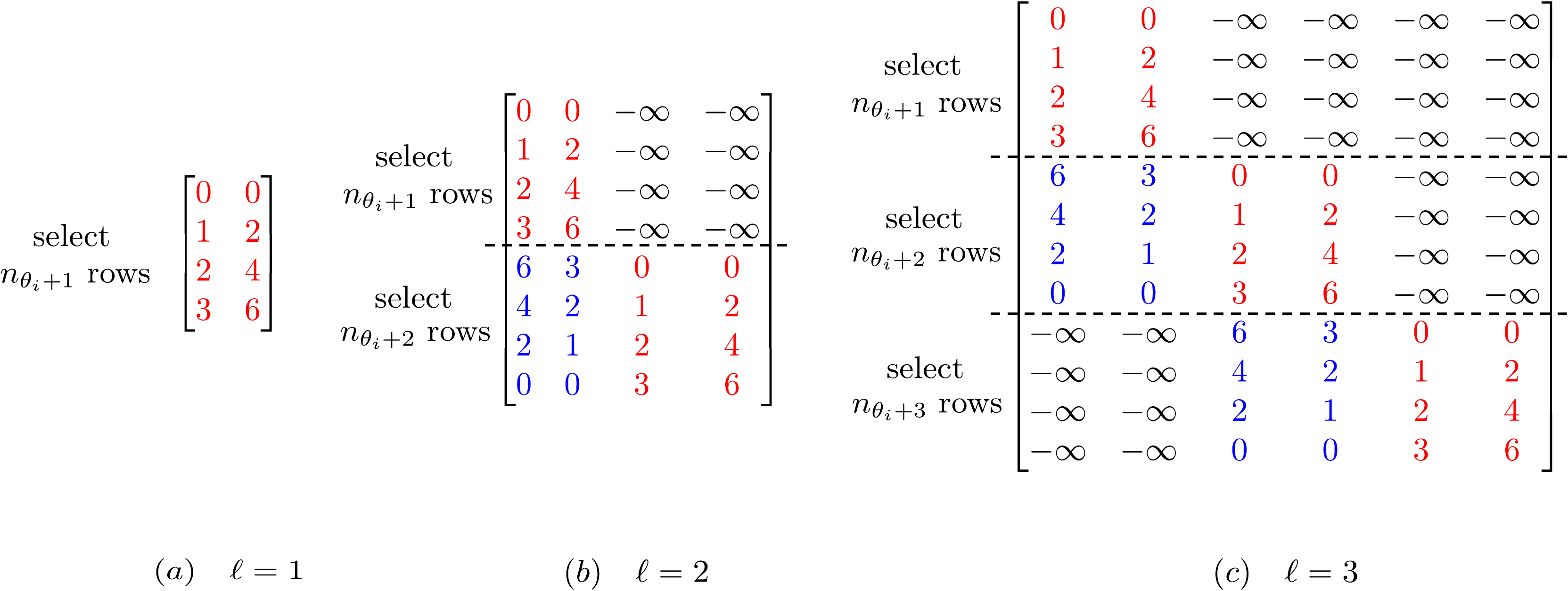}
		\caption{Let $\{n_{\theta_i+1},\ldots,n_{\theta_i+\ell}\}$ be such that they satisfy conditions (1)--(3) listed in Sec.~\ref{sec:example}. We will show that any $2\ell\times 2\ell$ exponent matrix $M_\text{dec}$ obtained by selecting rows as indicated above possesses a dominant permutation. }
		\label{fig:example_ell_1_2_3_exp}
	\end{center}
\end{figure*}

Towards arguing that all the exponent matrices possess dominant permutations, we first make the following remarks on $M^{(0)}$ and $M^{(1)}$.
\bit
\item[\bf{P1}] For each column, the elements of $M^{(0)}$ are strictly increasing from top to bottom.  
\item[\bf{P2}] For each column, the elements of $M^{(1)}$ are strictly increasing from bottom to top.
\item[\bf{P3}] For $r\in[0:2]$, any $2 \times 2$ matrix formed by stacking $r$ distinct rows of $M^{(0)}$ followed by $2-r$ distinct rows of $M^{(1)}$ has a dominant permutation.
\eit

{\it \underline{Case $\ell=1$}}: Here, there is just one choice for $(n_{\theta_i+1})$ (satisfying conditions (1)--(3)), which is $n_{\theta_{i}+1}=2$. As can be noted from Fig.~\ref{fig:example_ell_1_2_3_exp}(a), the exponent matrix $M_\text{dec}$ in this case is obtained by selecting $2$ rows of $M^{(0)}$. From {\bf P3} (with the choice $r=2$), it follows that any $2\times 2$ submatrix of $M^{(0)}$ possesses a dominant permutation.

{\it \underline{Case $\ell=2$}}: There are two possibilities for $(n_{\theta_i+1},n_{\theta_i+2})$ here; $(0,4)$ and $(1,3)$. In Fig.~\ref{fig:example_l_2_proof}, we illustrate these two scenarios. In the figure, we use red-colored and blue-colored asterisks as placeholders for the 
elements $M^{(0)}{(i,j)}$'s and $M^{(1)}{(i,j)}$'s, respectively. We partition the columns of these $4\times 4$ matrices into two thick columns. For $i \in [\ell]$, thick column $i$ consists of columns $[(i-1)k+1:ik]$.

\begin{figure}[!]
	\begin{center}
		\includegraphics[scale=0.5]{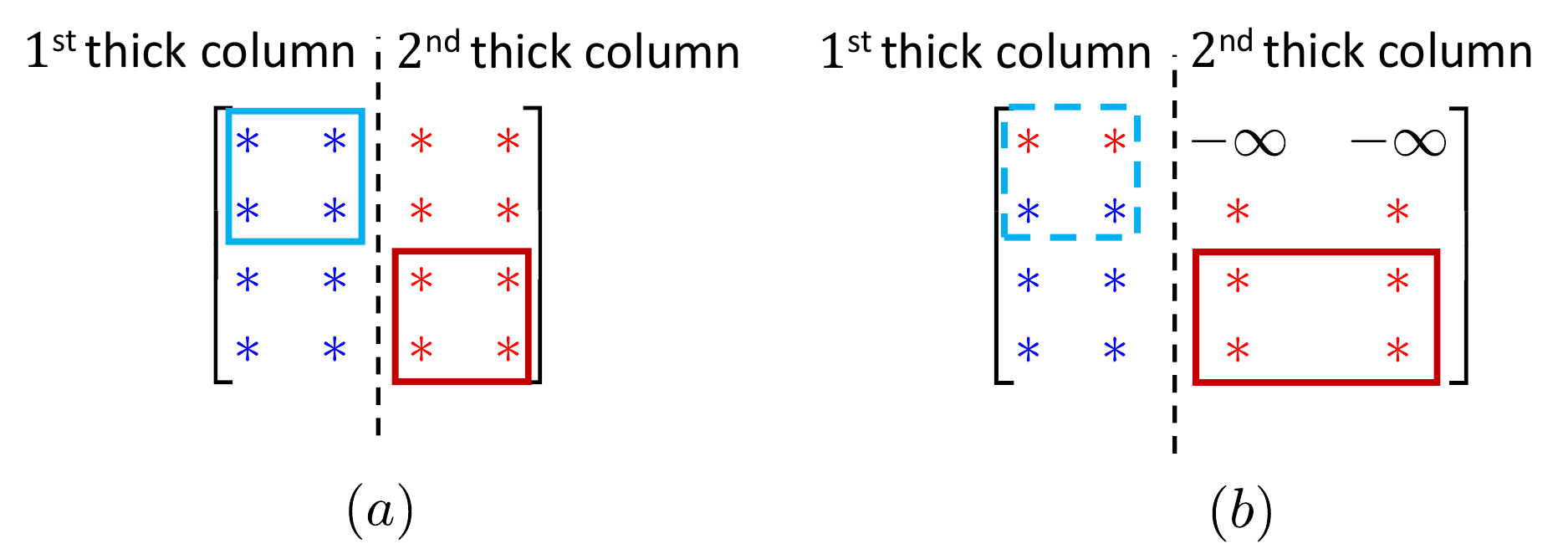}
		\caption{Here, we consider $\ell=2$. (a) $(n_{\theta_i+1},n_{\theta_i+2})=(0,4)$; (b) $(n_{\theta_i+1},n_{\theta_i+2})=(1,3)$. Vertical dashed lines demarcate thick columns. Submatrices demarcated using solid red and blue rectangles are dominant submatrices. The constrained dominant submatrix is indicated using a blue dashed rectangle.}
		\label{fig:example_l_2_proof}
	\end{center}
\end{figure} 

For the $(0,4)$ scenario, by {\bf P1} and {\bf P3} (with $r=2$), the lowermost $2\times 2$ submatrix (demarcated by a red rectangle in Fig.~\ref{fig:example_l_2_proof}(a)) is the dominant submatrix of the $2^\text{nd}$ thick column. Similarly, the topmost $2\times 2$ submatrix (shown by a blue rectangle) is the dominant submatrix of the $1^\text{st}$ thick column by {\bf P2} and {\bf P3} (with $r=0$). Because the rows of these submatrices do not overlap, any $4\times 4$ exponent matrix $M_\text{dec}$ of the form illustrated in  Fig.~\ref{fig:example_l_2_proof}(a) has a dominant permutation using the submatrix decomposition approach.

For the $(1,3)$ scenario, using similar arguments, the lowermost $2\times 2$ submatrix (demarcated by a red rectangle in Fig.~\ref{fig:example_l_2_proof}(b)) may be shown to be the dominant submatrix of the $2^\text{nd}$ thick column. As $M_\text{dec}(1,3)=M_\text{dec}(1,4)=-\infty$, if there is a dominant permutation $\sigma^*$, it should be that $1\in\sigma^*(\{1,2\})$ (similar to the scenario in Example \ref{ex:constrained_dom}). As a result, for the $1^\text{st}$ thick column, we limit our focus to those $2\times 2$ submatrices that involve row $1$. It is worth noting that row $1$ of the $1^\text{st}$ thick column is a row of $M^{(0)}$, whereas any other row of the thick column is a row of $M^{(1)}$. If we augment one row each from $M^{(0)}$ and $M^{(1)}$, the property {\bf P3} (with $r=1$) ensures the presence of a dominant permutation. Furthermore, based on {\bf P2}, the topmost $2\times 2$ submatrix (shown using a blue dashed rectangle) is the (constrained) dominant submatrix of the $1^\text{st}$ thick column. Because the rows of the red and blue dominant submatrices do not intersect, using submatrix decomposition, it follows that any $4\times 4$ exponent matrix $M_\text{dec}$ of the form illustrated in  Fig.~\ref{fig:example_l_2_proof}(b) has a dominant permutation.

{\it \underline{Case $\ell=3$}}: There are four possibilities for $(n_{\theta_i+1},n_{\theta_i+2},n_{\theta_i+3})$ here; $(0,2,4)$, $(0,3,3)$, $(1,1,4)$ and $(1,2,3)$. We illustrate these four cases in Fig.~\ref{fig:example_l_3_proof}. The basic idea remains the same as in the $\ell=2$ case, i.e., identification of dominant submatrices whose rows do not intersect. By {\bf P1} and {\bf P3} (with $r=2$), it follows that the submatrices demarcated by red rectangles within the $3^\text{rd}$ thick column are dominant submatrices (for all the four scenarios). For the $(0,2,4)$ scenario illustrated in Fig.~\ref{fig:example_l_3_proof}(a), due to the presence of $-\infty$ elements within the $1^\text{st}$ thick column of $M_\text{dec}$, it may be noted that any dominant permutation $\sigma^*$ (if exists) should satisfy $\sigma^*([3:6])=[3:6]$. As a result, we will search for the (constrained) dominant submatrix within the $2^\text{nd}$ thick column such that only rows $\{3,4,5,6\}$ are permitted. It now follows from properties {\bf P2} and {\bf P3} (with $r=0$) that the blue dashed rectangle depicts the constrained dominant submatrix of the $2^\text{nd}$ thick column. The submatrix highlighted in green is a dominant submatrix by {\bf P3} (with $r=0$). Because there is no intersection of rows among these dominant submatrices, the $6\times 6$ matrix has a dominant permutation.

\begin{figure*}[ht!]
	\begin{center}
		\includegraphics[scale=0.5]{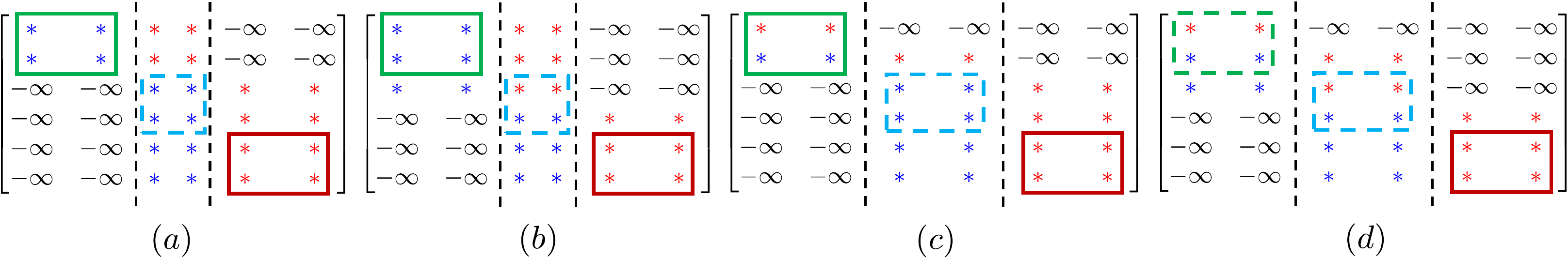}
		\caption{Here, we consider $\ell=3$. There are four possibilities for $(n_{\theta_i+1},n_{\theta_i+2},n_{\theta_i+3})$; (a) $(0,2,4)$, (b) $(0,3,3)$, (c) $(1,1,4)$ and (d) $(1,2,3)$. Vertical dashed lines demarcate thick columns. Submatrices demarcated using solid red and green rectangles are dominant submatrices. Constrained dominant submatrices are delineated using blue and green dashed rectangles.}
		\label{fig:example_l_3_proof}
	\end{center}
\end{figure*}

For the $(1,2,3)$ scenario (illustrated in Fig.~\ref{fig:example_l_3_proof}(d)), due to the $-\infty$'s in the $1^\text{st}$ thick column, it should be that $\sigma^*([3:6])\supseteq[4:6]$. Because of the $-\infty$'s in the $3^\text{rd}$ thick column, it should be that $\sigma^*([5:6])\subseteq[4:6]$. It follows that $\sigma^*([3:4])$ should contain precisely one element from $[4:6]$. Moreover, due to the $-\infty$'s in the $2^\text{nd}$ thick column, we have: $\sigma^*([3:4])\subseteq [2:6]$. Thus, $\sigma^*([3:4])$ should contain precisely one element from $[2:3]$. Pictorially, these constraints correspond to selecting one blue row and one red row within the $2^\text{nd}$ thick column. Due to {\bf P1}, {\bf P2} and {\bf P3} (with $r=1$), it follows that the submatrix demarcated using a blue dashed rectangle is the constrained dominant submatrix.  With regard to the $1^\text{st}$ thick column, owing to the constraint that $1 \notin \sigma^*([3:6])$, we have that $1\in\sigma^*([1:2])$ and  $|\sigma^*([1:2])\cap[2:3]|=1$. This corresponds to selecting the red row and one of the two blue rows in the $1^\text{st}$ thick column. From {\bf P2} and {\bf P3} (with $r=1$), it follows that the submatrix indicated in the green dashed rectangle is the constrained dominant submatrix. As rows of these three submatrices do not intersect, it follows that the $6\times 6$ matrix possesses a dominant permutation. Proofs for the scenarios $(0,3,3)$ and $(1,1,4)$ follow along similar lines. Since the largest term of matrices $M^{(0)}$, $M^{(1)}$ is $6$, it can be noted that the dominant sum of any exponent matrix will be at most $6*k*(\tau+1)=36$. Hence, choosing $d>36$ guarantees that all the decoding matrices are non-singular (by Lemma~\ref{lem:dominant_perm_non_singularity}).

\subsection{Construction A for any $\{n,k,m=1,\tau\}$}\label{sec:explicit_unit_mem_constr}

In this subsection, we describe our unit memory, explicit construction for any $\{n,k,\tau\}$.

\bconstr
\label{constr:explicit_unit_mem}\normalfont
Let $\alpha$ be a primitive element of $\mathbb{F}_{2^d}$. We describe an $(n, k, m=1)$ convolutional code over $\mathbb{F}_{2^d}$ by defining the $(n \times 2k)$ generator matrix $G = [G^{(1)} \ G^{(0)}]$, where $G^{(0)} = \alpha^{M^{(0)}}, G^{(1)} = \alpha^{M^{(1)}}$. We choose the $n \times k$ matrices $M^{(0)}, M^{(1)}$ as follows:
\bean
M^{(0)}(i,j) = (i-1)*j,  \ \ M^{(1)}(i,j) = (n-i)*(k+1-j). 
\eean
\econstr
\bthm \label{thm:constrA}
The $(n,k,m=1)$ convolutional code defined by Construction A is an $(n,k,m=1,\tau)$ $i$DOS code over $\mathbb{F}_{2^d}$ if $d>(n-1)k^2(\tau+1)$.
\ethm
The proof of the theorem is in Appendix~\ref{sec:thm_constrA_proof}.  

\section{Construction B: Explicit  Construction For All Parameters} \label{sec:general_construction}
The following convolutional code construction is a special case of the construction in \cite{AlmNapPin}. We will show that this code is an $(n,k,m,\tau)$  $i$DOS code  if the finite field size is sufficiently large.    
\bconstr\label{constr:explicit_all}\normalfont
Let $\alpha$ be a primitive element of $\mathbb{F}_{2^d}$. We describe an $(n, k, m)$ convolutional code over $\mathbb{F}_{2^d}$ by defining the $(n \times (m+1)k)$ generator matrix $G = [G^{(m)} \ \cdots \ G^{(0)}]$. For $t \in [0:m]$, $G^{(t)} \in \mathbb{F}_{2^d}^{n \times k}$  takes the form $G^{(t)} = \alpha^{M^{(t)}}$. We choose the $n \times k$ matrices $\{M^{(t)}\}$ as follows:
\bean 
M^{(t)}(i,j) = 2^{t*n+i+k-1-j}.
\eean
\econstr

\begin{example}\normalfont\label{sec:constrB_example}
The generator matrix as per Construction \ref{constr:explicit_all} for parameters $(n = 4, k = 2, m = 2)$ is given by: $$G = [\alpha^{M^{(2)}} \ \alpha^{M^{(1)}} \ \alpha^{M^{(0)}}],$$ where:
\bean 
M^{(0)} = \left[\begin{array}{cc}
	2^1 & 2^0\\
	2^2 & 2^1\\
	2^3 & 2^2\\
	2^4 & 2^3	
\end{array}\right], M^{(1)} = \left[\begin{array}{cc}
	2^5 & 2^4\\
	2^6 & 2^5\\
	2^7 & 2^6\\
	2^8 & 2^7	
\end{array}\right] \\ \text{ and } M^{(2)} = \left[\begin{array}{cc}
	2^9 & 2^8\\
	2^{10} & 2^9\\
	2^{11} & 2^{10}\\
	2^{12} & 2^{11}	
\end{array}\right].
\eean
\end{example}

\bthm\label{thm:explicit_all}
The $(n,k,m)$ convolutional code defined by Construction~\ref{constr:explicit_all} is an $(n, k, \tau, m)$ $i$DOS code over $\mathbb{F}_{2^d}$ if $ d>2^{((m+1)n+k-2)}(\tau+1) k$.  
\ethm

The proof can be found in Appendix~\ref{sec:explicit_all_proof}. 

\begin{remark}\normalfont
	When $m=1$, 
	 Construction~\ref{constr:explicit_all} has a much larger field extension degree requirement of $d>2^{(2n+k-2)}(\tau+1) k$ compared to the $d>(n-1)k^2(\tau+1)$ requirement of Construction~\ref{constr:explicit_unit_mem}.
\end{remark}
\newpage 
\bibliographystyle{IEEEtran}
\bibliography{Streaming_arXiv}

 \clearpage
\appendix
\subsection{Leibniz Formula}\label{sec:leibniz}
Given an $x\times x$ matrix $A$, the determinant $\det(A)$ can be computed as:
\beq\label{eq:leibniz}
\det(A)=\sum_{\sigma\in S_x}sgn(\sigma)\prod_{i=1}^x A({\sigma(i),i}).
\eeq
Here, $sgn(\sigma)=\pm 1$ ($+1$ if $\sigma$ is an even permutation, $-1$ if it is an odd permutation).

\subsection{Primitive Elements and Polynomials} \label{sec:prim_element_min_poly}

We quote the following well-known result as a lemma. 

\begin{lem}[{\cite[Ch.~4]{ecc_Mac_Slo}}]\label{lem:min_poly_lemma}
	Let $f(x)$ denote a non-zero polynomial of degree at most $d-1$ with coefficients over the subfield $\mathbb{F}_p$ (of $\mathbb{F}_{p^d}$). If $\alpha$ is a primitive element of $\mathbb{F}_{p^d}$, then $f(\alpha)\neq 0$.
\end{lem}

\subsection{Proof of Lemma~\ref{lem:dominant_perm_non_singularity}}
\label{sec:dominant_perm_non_singularity_proof}
\bpf
Let $A\triangleq \alpha^M$. For $\sigma\in S_x$, let $s_\sigma\triangleq \sum_{i=1}^x{M{(\sigma(i),i)}}$. From \eqref{eq:leibniz}, we have:
\bean
det(A)&=&\sum_{\sigma\in S_x}sgn(\sigma)\prod_{i=1}^x \alpha^{M{(\sigma(i),i)}}\\
&=& \sum_{\sigma\in S_x}sgn(\sigma) \alpha^{\sum_{i=1}^xM{(\sigma(i),i)}}\\
&=& sgn(\sigma^*)\alpha^{s_{\sigma^*}}+\sum_{\sigma\in S_x\setminus\{\sigma^*\}}sgn(\sigma) \alpha^{s_\sigma}\\
&=& f(\alpha),
\eean
where $f(x)\triangleq sgn(\sigma^*)x^{s_{\sigma^*}}+\sum_{\sigma\in S_x\setminus\{\sigma^*\}}sgn(\sigma) x^{s_\sigma}$. Clearly, $f(x)$ has $\{+1,-1\}$ as its coefficients, which are drawn from the subfield $\mathbb{F}_p$. Moreover, as $s_{\sigma^*}>s_\sigma$, $f(x)$ is a non-zero polynomial. The lemma now follows from the application of Lemma \ref{lem:min_poly_lemma}.
\epf

\subsection{Proof of Theorem~\ref{thm:constrA}} 
\label{sec:thm_constrA_proof} 

To prove that a convolutional code is an $i$DOS code, we have to show that all the message symbols in $\{\underline{s}(t)\mid t\in [\theta_i+1:\theta_{i+1}]\}$ can be recovered by the receiver using the available non-erased code symbols $\{c_j(t)\mid t\in [\theta_i+1:\theta_{i+1}],j\in \mathcal{R}_t\}$, under the assumption that the message symbols in $\{\underline{s}(t')\mid t'\in [\theta_i]\}$ are already known to the receiver. Since the code is time-invariant, we can assume without loss of generality that $\theta_i=0$ and $\theta_{i+1}=\ell$.

Properties (1)--(3) noted earlier in Sec.~\ref{sec:example} can be generalized to the following:
\bit
\item[(1)] $\ell\leq \tau+1$,
\item[(2)] $\sum_{i=1}^\ell n_i =k\ell$,
\item[(3)]  $k(\ell'-1)\leq \sum_{i=1}^{\ell'}n_i<k\ell'$, $\ell'\in [\ell-1] $.
\eit
In other words, $\{n_1,\dots,n_{\ell}\}$ corresponding to any $(n,k,m=1,\tau)$-acceptable symbol erasure pattern  will satisfy these three properties. 
Let $\mathcal{R}_t=\{i_1,i_2,\ldots,i_{n_t}\}\subseteq[n]$ and $\underline{\tilde{c}}(t)\triangleq [c_{i_1}(t)~\cdots~c_{i_{n_t}}(t)]^T$.
At time $\ell$,  the decoder has to solve  the following matrix equation:
\beq \label{eq:G_dec}
\begin{bmatrix}
	\underline{\tilde{c}}(1)\\ \vdots \\ \underline{\tilde{c}}(\ell)
\end{bmatrix} =
G_\text{dec}\begin{bmatrix}
	\underline{s}(1)\\ \vdots \\ \underline{s}(\ell)
\end{bmatrix},
\eeq
where $G_\text{dec}$ is a {\it decoding matrix} of size $k\ell \times k\ell$. Note that $G_\text{dec}$ is dependent on the symbol erasure pattern. It is easy to see that the code is an $i$DOS code if $G_\text{dec}$ is non-singular for any symbol erasure pattern such that  $\{n_{1},\ldots,n_{\ell}\}$ satisfy the properties (1)--(3). 
In our construction, for each decoding matrix $G_\text{dec}$, we have a corresponding $k\ell \times k\ell$ {\it exponent matrix} $M_\text{dec}$ such that $G_\text{dec}=\alpha^{M_\text{dec}}$.

We first start by  presenting Lemma~\ref{lem:n_properties} that shows some properties that are satisfied by $\{n_1, \cdots, n_{\ell}\}$. These properties will be used in Lemmas~\ref{lem:count_lemma_1} and \ref{lem:count_lemma_2} to prove the constraints that need to be satisfied by the dominant permutation (if it exists) of the exponent matrix $M_\text{dec}$.

\begin{lem}\label{lem:n_properties}
	The sequence $\{n_1,n_2,\ldots,n_\ell\}$ produced by any $(n,k,m=1,\tau)$-acceptable erasure pattern should
	satisfy the following conditions:
	\bit
	\item[(a)] $n_\ell \geq k$,
	\item[(b)] $n_{\ell'}\neq 0$ for $\ell' \in [2:\ell]$,
	\item[(c)] $\sum_{i=\ell-\mu+2}^\ell n_i \leq \mu k < \sum_{i=\ell-\mu+1}^\ell n_i$, $\mu\in[\ell-1]$,
	\item[(d)] $n_{\ell'} + n_{\ell'+1} \ge k$ for any $\ell' \in [\ell-1]$.
	\eit
\end{lem}
\bpf Recall properties (1)--(3) defined in Appendix~\ref{sec:thm_constrA_proof}. 
\bit
\item[(a)] If $\ell=1$, from property (2), we have $n_\ell=k$. If $\ell\geq 2$, from property (3) (with the choice $\ell'=\ell-1$), we have $\sum_{i=1}^{\ell-1}n_i<k(\ell-1)$. Using (2), it follows that $n_\ell>k$.

\item[(b)] Since $n_\ell>k$, the statement is true for $\ell'=\ell$. We will now assume $\ell' \in [2:\ell-1]$. From property (3), we have $\sum_{i=1}^{\ell'-1}n_i<k(\ell'-1)$. If $n_{\ell'}=0$, we can then write $\sum_{i=1}^{\ell'} n_i <k(\ell'-1)$ resulting in a contradiction of the first part of property (3).

\item[(c)] From property (2) and the first part of property (3) (with $\mu= \ell-\ell'$), we have:
$\sum_{i=1}^\ell n_i-\sum_{i=1}^{\ell'}n_i>(\ell-\ell')k$. Thus, we have $\sum_{i=\ell-\mu+1}^\ell n_i>\mu k$, where $\mu\in[1:\ell-1]$. 

From property (2) and the second part of property (3) (with $\mu=\ell-\ell'+1$), we have:
$\sum_{i=1}^\ell n_i-\sum_{i=1}^{\ell-\mu+1}n_i\leq (\ell-\ell'+1)k$. Thus, we have $\sum_{i=\ell-\mu+2}^\ell n_i\leq \mu k$, where $\mu\in[2:\ell]$. If $\mu=1$, we have $0\triangleq \sum_{i=\ell+1}^\ell n_i\leq k$, which is trivially true.
\item[(d)] From property (2) and first part of property (3), it follows that for all $\ell' \in [\ell-1]$:
\bean
n_{\ell'} + n_{\ell'+1} &\ge& k\ell' - \sum\limits_{i=1}^{\ell-1}n_i \\
&\ge& k\ell' -k(\ell'-1) = k,
\eean
where the last inequality follows from the second part of property (3).
\eit
\epf
\begin{figure}[ht!]
	\begin{center}
		\includegraphics[scale=0.4]{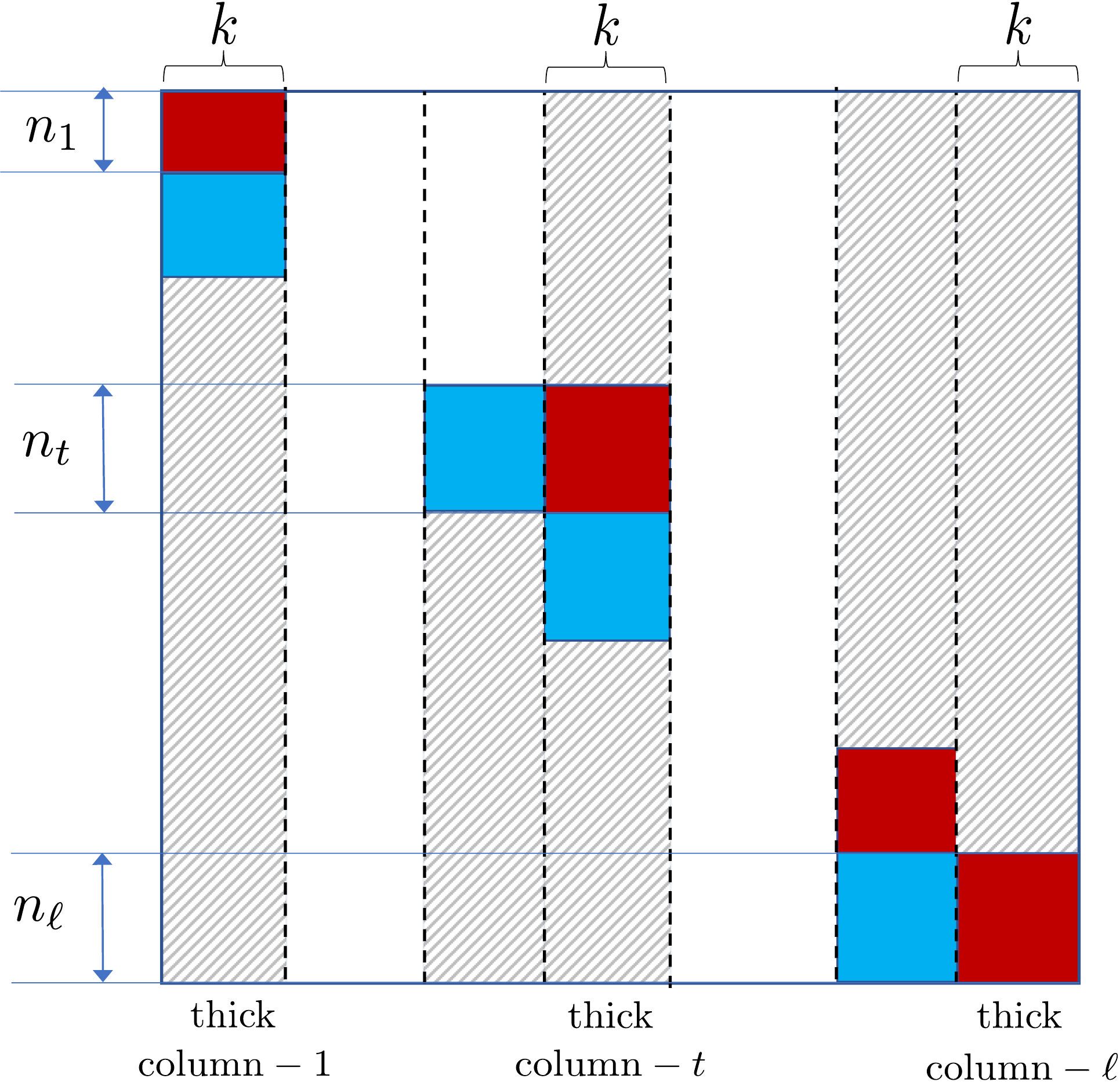}
		\caption{In this figure, we illustrate how the columns of the $k\ell \times  k\ell$ exponent matrix $M_\text{dec}$ are grouped into $\ell$ thick columns. Regions marked red and blue indicate submatrices of $M^{(0)}$ and $M^{(1)}$, respectively. Here, diagonal striped areas indicate $-\infty$ entries.}
		\label{fig:thick_cols}
	\end{center}
\end{figure}

From  Lemma \ref{lem:n_properties}(b), it follows that a given $k\ell \times k\ell$ exponent matrix $M_\text{dec}$, we can partition its columns into $\ell$ thick columns (see Fig.~\ref{fig:thick_cols}) where $\ell'$-th thick column has columns with indices ${\cal A}_{\ell'} = [(\ell'-1)k+1: \ell'k]$. Notice that for a permutation $\sigma^*$ to be dominant, for each thick column $\ell'$, the choice of the $k$ rows that get picked (given by $\sigma^*({\cal A}_{\ell'})$) has to come from the $n_{\ell'}+n_{\ell'+1}$ rows indicated by red and blue portions in Fig.~\ref{fig:thick_col_condition}.

In the upcoming Lemmas~\ref{lem:count_lemma_1} and \ref{lem:count_lemma_2}, we will further show the exact split of the red and blue rows that need to be picked by the dominant permutation for any thick column. 

\begin{lem}\label{lem:count_lemma_1}
	Suppose there exists a dominant permutation $\sigma^*$ for $M_\text{dec}$. Then for any thick column $(\ell-\mu)$ such that $\mu \in [1:\ell-1]$, $\sigma^*$ should satisfy that
	\bean
	|\sigma^*({\cal A}_{\ell-\mu}) \cap \hat{\cal B}_{\ell-\mu}| &=& r_{\mu} \text{ and ,}\\
	|\sigma^*({\cal A}_{\ell-\mu}) \cap \hat{\cal B}_{\ell-\mu+1}| &=& k-r_{\mu},
	\eean
	where $\hat{\cal B}_{\ell-\mu} = [\sum_{i=1}^{\ell-\mu-1}n_i:\sum_{i=1}^{\ell-\mu}n_i]$ and $r_{\mu}\triangleq (\mu+1)k-\sum_{i=\ell-\mu+1}^\ell n_i$. (See Fig.~\ref{fig:thick_col_condition} for an illustration).

	\begin{figure}
		\begin{center}
			\includegraphics[scale=0.43]{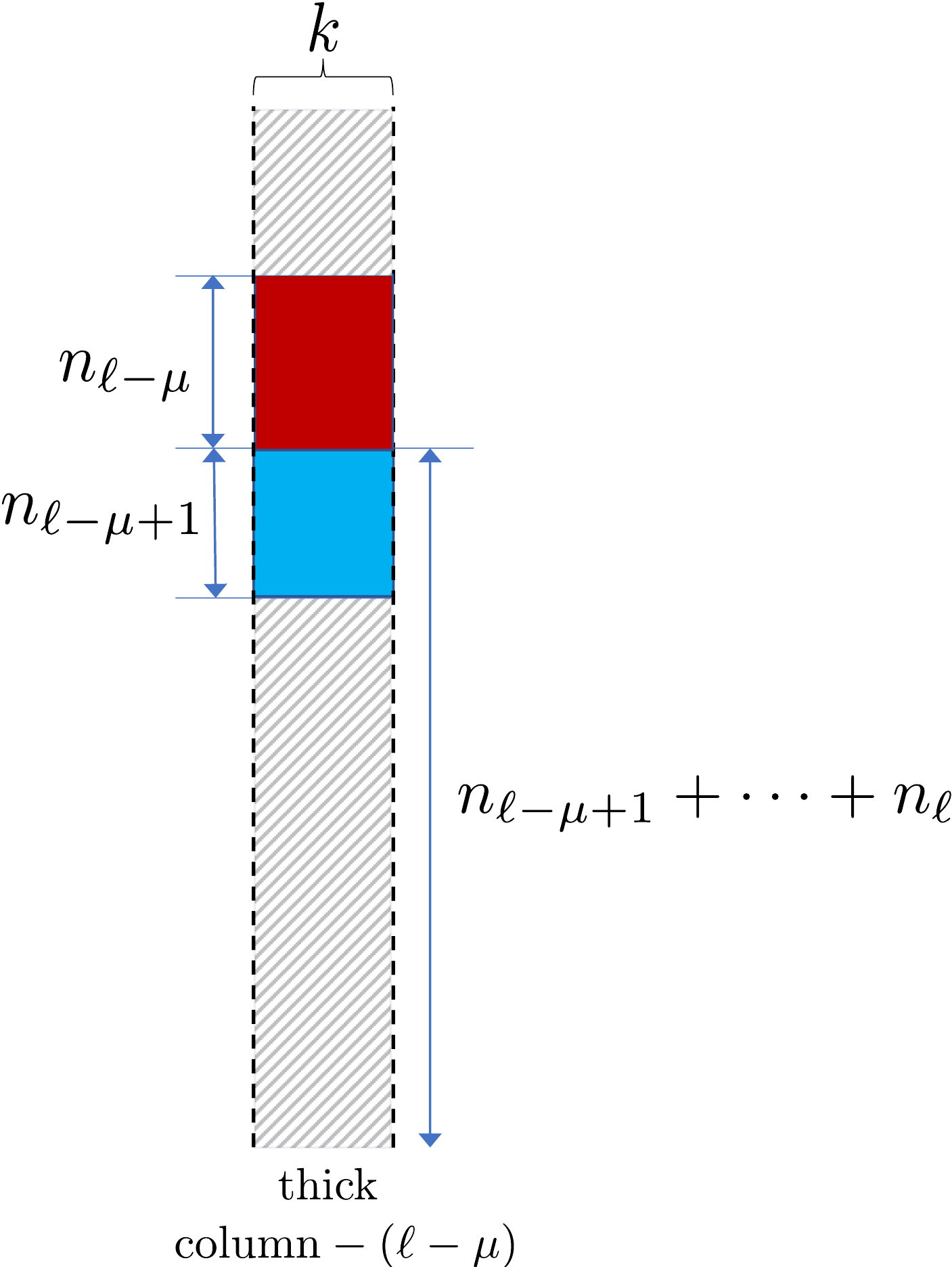}
			\caption{Let $\mu\in[\ell-1]$. Here, diagonal striped areas indicate $-\infty$ entries and regions marked red and blue indicate submatrices of $M^{(0)}$ and $M^{(1)}$, respectively and rows $\hat{\cal B}_{\ell-\mu}$, $\hat{\cal B}_{\ell-\mu+1}$ respectively. Suppose a dominant permutation $\sigma^*$ exists for $M_\text{dec}$. Consider the restriction of $\sigma^*$ to the thick column $(\ell-\mu)$, denoted by $\tilde{\sigma}^*_\mu$. Then, $\tilde{\sigma}^*_\mu$ involves precisely $(k-r_{\mu})$ blue rows and $r_{\mu}$ red rows. }
			\label{fig:thick_col_condition}
		\end{center}
	\end{figure}
\end{lem}

\bpf

Consider the Fig.~\ref{fig:general_proof_case_1}. 
Note that $\sigma^*({\cal A}_{\ell-\mu}) \subseteq \hat{\cal B}_{\ell-\mu} \cup \hat{\cal B}_{\ell-\mu+1}$ as the remaining rows see $-\infty$'s in the thick column given by ${\cal A}_{\ell -\mu}$.

\emph{Condition 1:} We first note that when restricted to thick columns $[\ell-\mu-1]$, the rows $\cup_{i=\ell-\mu+1}^{\ell} \hat{\cal B}_i $ 
consist of $-\infty$'s. As a result these rows need to be exhausted by the thick columns $\ell-\mu$ and beyond, i.e., 
\bean
\cup_{i=\ell-\mu+1}^{\ell} \hat{\cal B}_i \subseteq \sigma^*(\cup_{i=\ell-\mu}^{\ell} {\cal A}_i).
\eean
\emph{Condition 2:} Since the thick columns $[\ell-\mu+1:\ell]$, given by indices $\cup_{i=\ell-\mu+1}^{\ell}{\cal A}_i$ when restricted to the rows $\cup_{i=1}^{\ell-\mu}\hat{{\cal B}}_i$ consist of $-\infty$ it follows that:
\bean
\sigma^*(\cup_{i=\ell-\mu+1}^{\ell}{\cal A}_i) \subseteq \cup_{i=\ell-\mu+1}^{\ell} \hat{\cal B}_i.
\eean
Therefore, $\mu k$ rows out of the rows $\cup_{i=\ell-\mu+1}^{\ell}\hat{{\cal B}}_i$ are already used up by thick columns $\ell-\mu+1$ and beyond. The remaining rows $ \sum\limits_{i=\ell-\mu+1}^{\ell} n_i - \mu k = (k-r_{\mu})$ need to be exhausted by the thick column $\ell-\mu$ due to Condition 1 i.e.,
\bean
|\sigma^*({\cal A}_{\ell-\mu}) \cap \cup_{i = \ell-\mu+1}^{\ell}\hat{\cal B}_i| = (k - r_{\mu}).
\eean

Note that $r_{\mu}$ is well defined within $[0:k]$ due to Lemma~\ref{lem:n_properties}. Also from the observation that  $\sigma^*({\cal A}_{\ell-\mu}) \subseteq \hat{\cal B}_{\ell-\mu} \cup \hat{\cal B}_{\ell-\mu+1}$ it follows that
\bean
|\sigma^*({\cal A}_{\ell-\mu}) \cap \cup_{i = \ell-\mu+1}^{\ell}\hat{\cal B}_i| &=& |\sigma^*({\cal A})_{\ell-\mu} \cap \hat{\cal B}_{\ell-\mu+1}| \\&=& (k-r_{\mu}) \text{ and }\\
|\sigma^*({\cal A})_{\ell-\mu} \cap \hat{\cal B}_{\ell-\mu}| &=& r_{\mu}.
\eean

\begin{figure*}
	\begin{center}
		\includegraphics[scale=0.43]{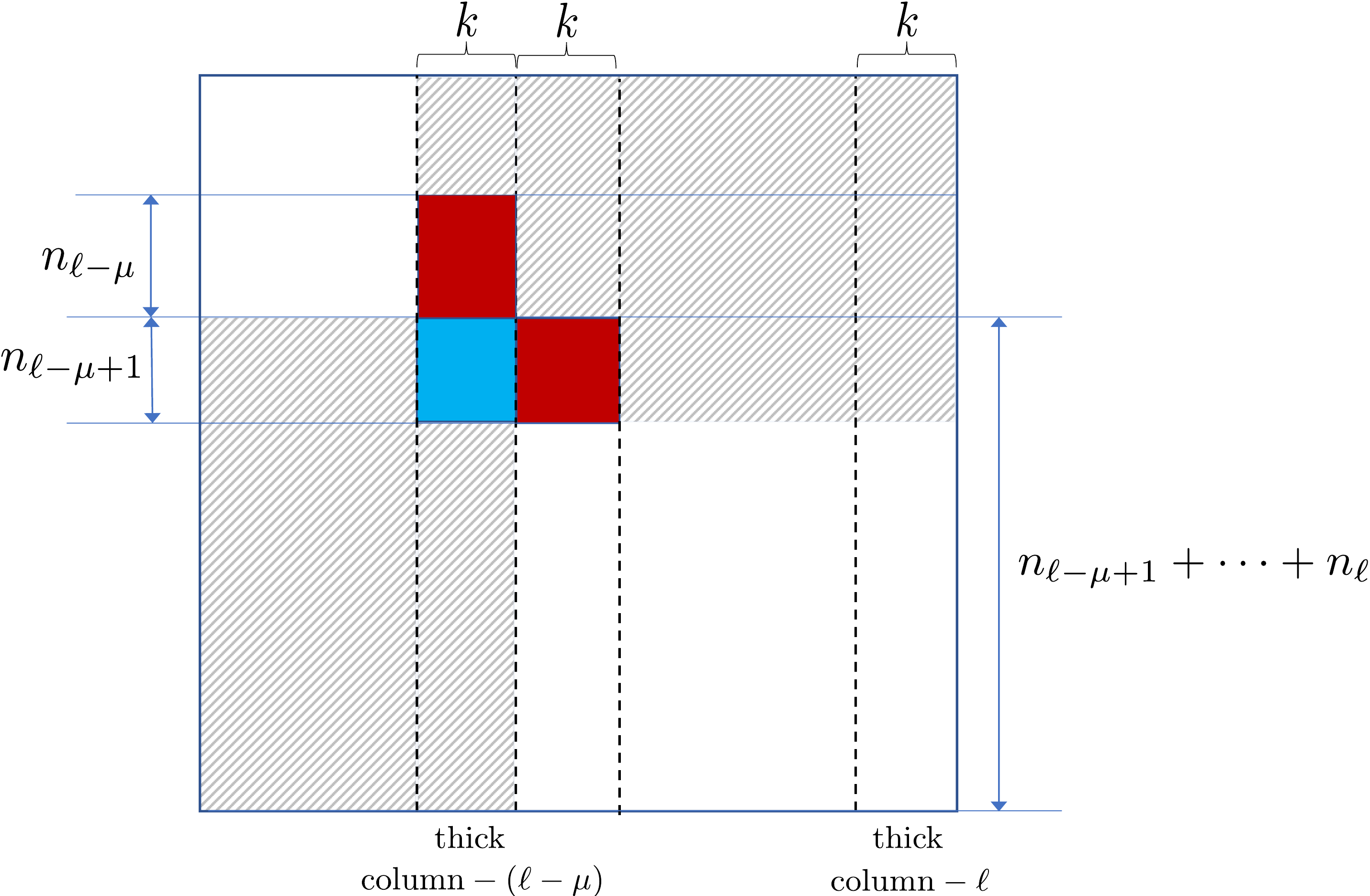}
		\caption{Figure highlighting structure of $M_\text{dec}$ matrix. Regions marked red and blue indicate submatrices of $M^{(0)}$ and $M^{(1)}$, respectively. Diagonal stripes indicate regions where we have $-\infty$'s. We do not care about the values of the unshaded regions.}
		\label{fig:general_proof_case_1}
	\end{center}
\end{figure*}
\epf

\begin{lem}\label{lem:count_lemma_2}
	Suppose there exists a dominant permutation $\sigma^*$ for $M_\text{dec}$ then 
	\bean
	|\sigma({\cal A}_{\ell}) \cap \hat{\cal B}_{\ell}| = k \triangleq r_0, 
	\eean
	where $\hat{\cal B}_{\ell}=[\ell k -n_{\ell}+1: \ell k]$. 
\end{lem}
\bpf
Clearly, $|\sigma({\cal A}_{\ell})|$ is equal to $k$. In the thick column $\ell$, all rows except the rows given by $\hat{ \cal B}_{\ell}$ consist of $-\infty$ elements. Thus, the lemma follows. 	
\epf

We will now examine a few properties of the matrices $M(0)$, $M(1)$ defined by Construction~\ref{constr:explicit_unit_mem}. Subsequently, we use these properties to prove Corollary~\ref{cor:M_dom_sub_mat}.

\blem\label{lem:M_properties}
The matrices $M^{(0)}, M^{(1)}$ defined in Construction~\ref{constr:explicit_unit_mem} satisfy the following three properties.
\bit
\item[\bf{P1}] For each column, the elements of $M^{(0)}$ are strictly increasing from top to bottom.  
\item[\bf{P2}] For each column, the elements of $M^{(1)}$ are strictly increasing from bottom to top.
\item[\bf{P3}] 
For $r\in[0:k]$, any $k \times k$ matrix formed by stacking $r$ distinct rows of $M^{(0)}$ followed by $2-r$ distinct rows of $M^{(1)}$ has a dominant permutation.
\eit
\elem
\bprf
The properties {\bf{P1}} and {\bf{P2}} follow directly from the definition of matrices $M^{(0)}, M^{(1)}$ defined in Construction~\ref{constr:explicit_unit_mem}.
Let $\hat{M}$ be the matrix obtained by augmenting $(k-r)$ rows indexed by $R_1 = \{i_1, \cdots, i_{k-r}\}$ of $M^{(1)}$ followed by $r$ rows $R_0 = \{i_{k-r+1}, \cdots, i_k\} $ of $M^{(0)}$. Then,
\bean
\hat{M} = \left[ \begin{array}{c}
	M^{(1)}(R_1,[k])\\
	M^{(0)}(R_0,[k])\\
\end{array}\right].
\eean
If $\hat{M}$ has a dominant permutation, then the $k \times k$ matrix considered in ${\bf{P3}}$ will also have a dominant permutation as these two matrices differ only in the ordering of rows. 
To prove property {\bf{P3}} we will show that the identity permutation $\sigma_0 \in S_k$ defined by $\sigma_0(i)=i$ for all $i \in [k]$ is the dominant permutation of the matrix  $\hat{M}$.

Let us consider any other permutation $\sigma \in S_k \setminus \{\sigma_0\}$; we will show that the sum corresponding to $\sigma$ is strictly less than the dominant sum. To do this, for any $\sigma \in S_k \setminus \{\sigma_0\}$, let $j$ be the smallest integer such that $\sigma(j) \ne \sigma_0(j) = j$, we will construct $\hat{\sigma}$ such that 
\ben
\item $\hat{\sigma}(i) = \sigma_0(i) = i$ for all $i \le j$ and
\item $\sum\limits_{i=1}^k \hat{M}{(\sigma(i),i)} < \sum\limits_{i=1}^k \hat{M}{(\hat{\sigma}(i),i)}$.
\een
We can then use this method to recursively construct a permutation that has a larger sum until we arrive at the  permutation $\sigma_0$, indicating that $\sigma_0$ is the dominant permutation.

Let $j'= \sigma^{-1}(j)$. We define the permutation $\hat{\sigma}$ as
\bean
\hat{\sigma} = \begin{cases}
	\sigma(i) & i \ne j, j' \\
	\sigma(j') = j & i = j\\
	\sigma(j) & i = j'.
\end{cases}
\eean
Note that $j' > j$ and $\sigma(j) > \sigma(j') = j$ by the definition of $j$ and $j'$. From the definition, it is also clear to see that $\hat{\sigma}(i) = i$ for all $i \le j$. We look at the difference between the sum corresponding to the permutation $\hat{\sigma}$ and the sum for the permutation $\sigma$ and prove that it is strictly positive.
\bean
\sum\limits_{i=1}^k \hat{M}{(\hat{\sigma}(i),i)} - \sum\limits_{i=1}^k \hat{M}{(\sigma(i),i)} = \hat{M}{(\hat{\sigma}(j),j)} + \hat{M}{(\hat{\sigma}(j'),j')} - \hat{M}{(\sigma(j),j)} - \hat{M}{(\sigma(j'),j')}.  
\eean
We divide the rest of the proof into three cases: (1) $j \le k-r, \sigma(j) \le k-r$, (2) $j \le k-r, \sigma(j) > k-r$ and (3) $j > k-r$ and argue for each case that the difference between sums above is strictly positive. 
\bit
\item[(1)] $j \le k-r, \sigma(j) \le k-r$: The difference between sums reduces to the following expression in this case:
\bean
 M^{(1)}(i_{\hat{\sigma}(j)},j) + M^{(1)}(i_{\hat{\sigma}(j')},j') - M^{(1)}(i_{\sigma(j)},j) - M^{(1)}(i_{\sigma(j')},j') = (i_{\sigma(j)}-i_{\sigma(j')})(j'-j) > 0.
\eean
\item[(2)] $j \le k-r, \sigma(j) > k-r$: The difference between sum reduces to the following expression:
\bean
M^{(1)}(i_{\hat{\sigma}(j)},j) + M^{(0)}(i_{\hat{\sigma}(j')},j') - M^{(0)}(i_{\sigma(j)},j) - M^{(1)}(i_{\sigma(j')},j') = (k-1+i_{\sigma(j)}-i_{\sigma(j')})(j'-j) > 0.
\eean
\item[(3)] $j > k-r$: The difference between sums reduces to the following expression:
\bean
&& M^{(0)}(i_{\hat{\sigma}(j)},j) + M^{(0)}(i_{\hat{\sigma}(j')},j') - M^{(0)}(i_{\sigma(j)},j) - M^{(0)}(i_{\sigma(j')},j') =  (i_{\sigma(j)}-i_{\sigma(j')})(j'-j) > 0.
\eean
\eit
\eproof

By applying Lemma~\ref{lem:M_properties},  we obtain the following result on the structure of constrained dominant submatrices.

\bcor\label{cor:M_dom_sub_mat}  
Let $r \in [0:k]$, $w_0, w_1 \in [0:n]$ such that $w_0 \ge r, w_1 \ge k-r$ and $\hat{M}$ be a $((w_0+w_1) \times k)$ matrix formed by augmenting the $w_0$ rows of $M^{(0)}$ followed by $w_1$ rows of $M^{(1)}$. Among the $(k \times k)$ submatrices of $\hat{M}$ that pick first $r$ rows from $M^{(0)}$ and later $(k-r)$ rows from $M^{(1)}$, 
the submatrix $\hat{M}([w_0-r+1:w_0-r+k],:)$ has the largest dominant sum.
\ecor
\bpf
From property {\bf{P3}} shown in Lemma~\ref{lem:M_properties} it follows that any constrained submatrix $\hat{M}({\cal A}, :)$  has a dominant permutation if  $|{\cal A} \cap [w_0]| = r$ and $|{\cal A} \cap [w_0+1:w_0+w_1]| = k-r$.  The constraint comes from the fact that the first $r$ rows of the submatrix come from $M^{(0)}$ and the rest $(k-r)$ come from $M^{(1)}$.
From {\bf{P1}}, the elements in $M^{(0)}$ increase from top to bottom. Therefore it follows that for any ${\cal A}$ such that ${\cal A} \cap [w_0] \neq [w_0-r+1:w_0]$, the dominant sum of $\hat{M}({\cal A},:)$ is strictly smaller than $\hat{M}(\hat{\cal A},:)$ where $\hat{\cal A} = [w_0-r+1:w_0] \cup ({\cal A} \cap [w_0+1:w_0+w_1])$. Similarly from {\bf{P2}}, the elements in $M^{(1)}$ decrease from top to bottom. Therefore, it follows that for any ${\cal A}$ such that ${\cal A} \cap [w_0+1:w_0+w_1] \neq [w_0+1:w_0+k-r]$, the dominant sum of $\hat{M}({\cal A},:)$ is strictly smaller than $\hat{M}(\tilde{\cal A},:)$ where $\tilde{\cal A} = ({\cal A} \cap [w_0]) \cup [w_0+1:w_0+k-r]$.
Therefore, the dominant sum of submatrix $\hat{M}([w_0-r+1:w_0+k-r],:)$ is the largest indicating that it is the constrained dominant submatrix.
\epf

We now have all the results needed to prove the Theorem~\ref{thm:constrA}.   

\emph{Proof of Theorem~\ref{thm:constrA}:}
We will show that there exists a dominant permutation for $M_\text{dec}$, provided $\{n_{1},\ldots,n_{\ell}\}$ satisfy the properties (1)--(3). 
This will, in turn, prove that the corresponding $G_{dec}$  is invertible due to Lemma~\ref{lem:dominant_perm_non_singularity}, thereby proving the Theorem~\ref{thm:constrA}. 

\subsubsection{Constrained Dominant Submatrix of  Thick Column $(\ell-\mu)$ for $ \mu \in [0:\ell-1]$} From the Lemmas~\ref{lem:count_lemma_1} and \ref{lem:count_lemma_2}, if there exists a dominant permutation $\sigma^*$ of $M_\text{dec}$, for the $(\ell-\mu)$-th thick column it needs to pick $r_{\mu}$ rows from within indices $\hat{\cal B}_{\ell - \mu}$ and $(k-r_{\mu})$ rows from indices $\hat{\cal B}_{\ell - \mu+1}$. This is same as picking $r_{\mu}$ rows from matrix $M^{(0)}({\cal R}_{\ell -\mu},:)$ and $(k-\mu)$ rows from matrix $M^{(1)}({\cal R}_{\ell -\mu+1},:)$. See Fig.~\ref{fig:thick_col_condition} for an illustration.

From Corollary~\ref{cor:M_dom_sub_mat}, it follows picking the last $r_{\mu}$ rows of $M^{(0)}({\cal R}_{\ell-\mu}, :)$ out of the $n_{\ell-\mu}$ rows and the first $(k - r_{\mu})$ rows of $M^{(1)}({\cal R}_{\ell-\mu+1}, :)$ out of the $n_{\ell-\mu+1}$ rows results in the largest dominant sum. Therefore the rows of the constrained dominant submatrix that can be picked for the  thick column ${\cal A}_{\ell-\mu}$ are given by:
\bean
{\cal B}_{\ell-\mu} &=& [\sum\limits_{i=1}^{\ell-\mu}n_i - r_{\mu}+1: \sum\limits_{i=1}^{\ell-\mu}n_i-r_{\mu}+k]\\
&=& [(\ell-\mu-1)k+1:(\ell-\mu)k] = {\cal A}_{\ell-\mu}.
\eean
This indicates that the constrained dominant sub matrix of $\hat{M}(:,{\cal A}_{\ell-\mu})$ is $\hat{M}({\cal A}_{\ell-\mu},{\cal A}_{\ell-\mu})$.
\subsubsection{Disjoint Row Property} It is clear to see that the rows selected by the constrained dominant submatrices, ${\cal B}_{\ell-\mu} = {\cal A}_{\ell-\mu}$ are disjoint. Therefore, by the submatrix decomposition strategy  given in Remark~\ref{rem:dom_sub_matrix}, there exists a dominant permutation $\sigma^*$ and it is such that $\sigma^*({\cal A}_{\ell-\mu}) = {\cal A}_{\ell-\mu}$ for all $\mu \in [0:\ell-1]$.
\subsubsection{Field Size Requirement}Since the largest term in the matrices $M^{(0)}$, $M^{(1)}$ is $(n-1)k$, it can be noted that the dominant sum of any decoding matrix will be at most $(n-1)k^2(\tau+1)$.
Therefore, given $d> (n-1)k^2(\tau+1)$, by Lemma~\ref{lem:dominant_perm_non_singularity}, $G_{\text{dec}}$ is invertible for any $(n,k,m,\tau)$-acceptable erasure pattern. 
\eproof

\subsection{Superregular Matrix} \label{sec:superregular}
Each product of the form $\prod_{i=1}^x A({\sigma(i),i})$ in \eqref{eq:leibniz} corresponding to the permutation $\sigma$, is referred to as a {\it term}. A term is a trivial  term if $A({\sigma(i),i})=0$ for some $i \in [x]$. Consider a $y\times z$ matrix $B$. An $x\times x$ square submatrix $C$ of $B$ is said to be a non-trivial submatrix if at least one of the terms in the Leibniz formula for $\det(C)$ is non-trivial. The matrix $B$ is {\it superregular} if all its non-trivial submatrices are non-singular.

\subsection{Proof of Theorem~\ref{thm:explicit_all}}
\label{sec:explicit_all_proof}

We first restate a theorem from  \cite{AlmNapPin} that presents an explicit construction of superregular matrix. See Appendix~\ref{sec:superregular} for the definition of superregular matrices. 
\bthm\label{thm:superreg}
Let $\alpha$ be a primitive element of a finite field $\mathbb{F}_{p^{d}}$ and $B=[v_{i,j}]$ be a matrix over $\mathbb{F}_{p^{d}}$ with the following properties
\bit
\item if $v_{i,j} \ne 0$ then $v_{i,j} = \alpha^{\beta_{i,j}}$ for a positive integer $\beta_{i,j}$;
\item if $v_{i, j} = 0$ then $v_{i',j} = 0$ , for any $i' > i$ or $v_{i,j'} = 0$, for any 	$j' > j$;
\item if $j' < j$ and $v_{i,j} \ne 0$ and $v_{i,j'} \ne 0$ then $2\beta_{i,j} \le \beta_{i,j'}$.
\item if $i < i'$ and $v_{i,j} \ne 0$ and $v_{i',j} \ne 0$ then $2\beta_{i,j} \le \beta_{i',j}$.
\eit
Suppose $d$ is greater than any exponent of $\alpha$ appearing as a non-trivial term in the Leibniz formula for the determinant of any square submatrix of $B$. Then $B$ is superregular.
\ethm

\begin{figure}[ht!]
	\begin{center}
		\includegraphics[scale=0.42]{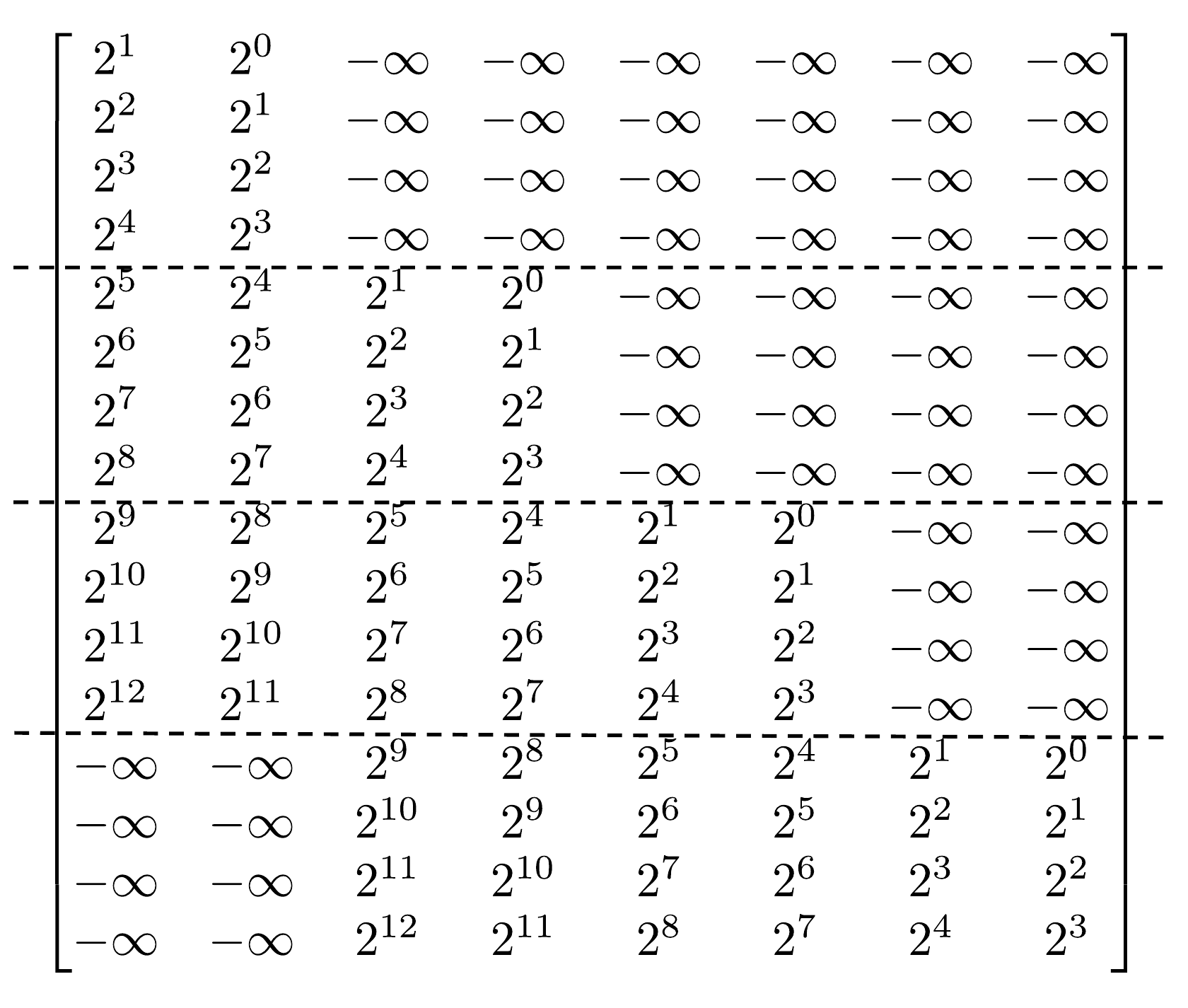}
		\caption{Let $\check{G}=\alpha^{\check{M}}$. For $(n=4,k=2,m=2)$ example construction given in Example~\ref{sec:constrB_example}, the figure shows the matrix $\check{M}$ for $\ell=4$ case.}
		\label{fig:general_constr_example}
	\end{center}
\end{figure}

\emph{Proof of Theorem~\ref{thm:explicit_all}:} As in the beginning of Appendix~\ref{sec:thm_constrA_proof}, we can assume without loss of generality that $\theta_i=0$ and $\theta_{i+1}=\ell$. 
Let  $\check{G}$ be the $n\ell \times k \ell$ matrix such  
\beqn
\begin{bmatrix}
	\underline{c}(1)\\ \vdots \\ \underline{c}(\ell)
\end{bmatrix} =
\check{G}\begin{bmatrix}
	\underline{s}(1)\\ \vdots \\ \underline{s}(\ell)
\end{bmatrix}. 
\eeqn

See Fig.~\ref{fig:general_constr_example} for an example illustration of $\check{G}$ matrix.   It can be verified that $\check{G}$ satisfies all the four conditions  stated in Theorem \ref{thm:superreg}. Now recall the definition of the $\ell k \times \ell k$ matrix $G_\text{dec}$ in \eqref{eq:G_dec}. It can be seen that $G_\text{dec}$ is a square submatrix of $\check{G}$. The code defined in Construction~\ref{constr:explicit_all} is an $(n,k,m,\tau)$ $i$DOS code if $G_\text{dec}$ is non-singular for any  $\{n_{1},\ldots,n_{\ell}\}$ corresponding to an $(n,k,m,\tau)$-acceptable symbol erasure pattern.
Since all $M^{(t)}(i,j) \le 2^{mn+n+k-2},$ the exponent of $\alpha$ appearing as a non-trivial term in the determinant of any square submatrix of $\check{G}$ is upper bounded by $2^{((m+1)n+k-2)}\ell k$.   If $\ell \le \tau+1$, $\check{G}$ is superregular due to Theorem~\ref{thm:superreg} and every non-trivial square submatrix of $\check{G}$  is non-singular. For all $(n,k,m,\tau)$-acceptable symbol erasure patterns, we have $\ell \le \tau+1$.
By superregular property of $\check{G}$, to show that $G_\text{dec}$ is non-singular, it suffices to show that the determinant of $G_\text{dec}$ has a non-trivial term. This follows from  \cite[Lemma~4]{idos}, in which it is shown that all the diagonal entries are non-zero for $G_\text{dec}$ of any $(n,k,m,\tau)$-acceptable symbol erasure pattern.  
\eproof
\end{document}